\documentclass[conference]{IEEEtran}

\usepackage{pdflscape}
\usepackage{booktabs}
\usepackage{array}
\usepackage{ragged2e}
\usepackage{adjustbox}
\usepackage{tabularx}
\usepackage[T1]{fontenc}
\usepackage[utf8]{inputenc}
\usepackage{longtable}
\usepackage{multirow}
\usepackage{textgreek}
\newcolumntype{L}[1]{>{\RaggedRight\arraybackslash}p{#1}}
\IEEEoverridecommandlockouts
\usepackage{cite}
\usepackage{amsmath,amssymb,amsfonts}
\usepackage{algorithmic}
\usepackage{textcomp}
\usepackage{subfigure}
\usepackage{xcolor}
\def\BibTeX{{\rm B\kern-.05em{\sc i\kern-.025em b}\kern-.08em T\kern-.1667em\lower.7ex\hbox{E}\kern-.125emX}}
\usepackage{array}
\usepackage[Symbol]{upgreek}
\usepackage{booktabs}
\usepackage{longtable}
\usepackage{mathtools}
\usepackage{capt-of}
\usepackage{tikz}
\usepackage{float}
\usetikzlibrary{tikzmark}
\usetikzlibrary{arrows.meta}
\usepackage{multirow}
\usepackage{longtable}
\usepackage{tabularx}
\usepackage{hyperref}
\hypersetup{colorlinks=true,linkcolor=blue,	citecolor=blue,}
\usepackage{dblfloatfix}
\usepackage{footnote}
\usepackage{makecell}
\usepackage{enumitem}
\usepackage{pifont}
\usepackage{forest,array}
\usepackage[utf8]{inputenc}
\usepackage{tabularray}
\usepackage{subcaption}
\usepackage{pdflscape}
\usepackage{lscape}
\usepackage{float}
\usepackage{rotating}
\usepackage{supertabular,booktabs}
\usetikzlibrary{shapes}
\usepackage{soul}
\definecolor{lightsalmonpink}{RGB}{255, 180, 190}
\definecolor{lavender}{RGB}{180, 165, 200}
\definecolor{lemonchiffon}{RGB}{255, 250, 205}  
\definecolor{salmonpink}{RGB}{255, 145, 164}
\sethlcolor{salmonpink}
\definecolor{verylightsalmonpink}{RGB}{255, 210, 215}
\usepackage[dvipsnames]{xcolor}
\usepackage{titlesec}
\usepackage{hyperref}
\usepackage[justification=centering]{caption}
\usepackage{diagbox}
\begin{document}
\setlength{\parskip}{0pt} 
\setlength{\parindent}{15pt}
\title{PTMF: A Privacy Threat Modeling Framework for IoT with Expert-Driven Threat Propagation Analysis  \\ \Large \color{red}\textbf{Authors’ draft for soliciting feedback}
\today \\}
\author{
    \textbf Emmanuel Alalade, Ashraf Matrawy\\ School of Information Technology, Carleton University, Canada \\ \\Emails: emmanuelalalade@cmail.carleton.ca
}
\maketitle
\pagestyle{plain}

\begin{abstract}

Previous studies on PTA have focused on analyzing privacy threats based on the potential areas of occurrence and their likelihood of occurrence. However, an in-depth understanding of the threat actors involved, their actions, and the intentions that result in privacy threats is essential. In this paper, we present a novel Privacy Threat Model Framework (PTMF) that analyzes privacy threats through different phases.

The PTMF development is motivated through the selected tactics from the MITRE ATT\&CK framework and techniques from the LINDDUN privacy threat model, making PTMF a privacy-centered framework. The proposed PTMF can be employed in various ways, including analyzing the activities of threat actors during privacy threats and assessing privacy risks in IoT systems, among others. In this paper, we conducted a user study on 12 privacy threats associated with IoT by developing a questionnaire based on PTMF and recruited experts from both industry and academia in the fields of security and privacy to gather their opinions. The collected data were analyzed and mapped to identify the threat actors involved in the identification of IoT users (IU) and the remaining 11 privacy threats. Our observation revealed the top three threat actors and the critical paths they used during the IU privacy threat, as well as the remaining 11 privacy threats. This study could provide a solid foundation for understanding how and where privacy measures can be proactively and effectively deployed in IoT systems to mitigate privacy threats based on the activities and intentions of threat actors within these systems.

\end{abstract}

 
\section{Introduction}

To implement a privacy measure in IoT systems, the first step is to conduct a Privacy Threat Analysis (PTA) to identify potential vulnerabilities and privacy threats that can exploit these vulnerabilities in the IoT system \cite{LINDDUN, LINDDUNTree}. 

The PTA process serves as a systematic approach for evaluating privacy threats, identifying potential risks associated with these threats, and informing the development of appropriate countermeasures. However, existing research on the PTA process in IoT systems primarily focuses on analyzing privacy threats based on where these threats can occur in IoT systems and which assets they could impact if they were to occur \cite{ninggal2011privacy,estrada2017online,jawurek2011privacy,iwaya2018mhealth}. These previous studies are limited because they do not provide sufficient information on the activities of threat actors and the tactical approaches these actors employed to achieve privacy threats. Consequently, there exists a compelling need to develop a more comprehensive privacy threat analysis framework that incorporates a detailed characterization of threat actors' activities, enabling the precise implementation of Privacy Preservation Techniques (PPTs) tailored to specific threat actors' tactics and techniques.

To address the limitations identified in previous studies on PTA methodologies, this paper presents a novel, comprehensive framework, the Privacy Threat Modeling Framework (PTMF), as shown in \autoref{fig:PTMF}. This framework is designed for various privacy preservation processes, including analyzing the activities of threat actors responsible for specific privacy threats in IoT systems. To empirically examine the application of the PTMF, we analyzed 12 privacy threats (that is, T1-T12) with a detailed analysis on  "T1-Identification of IoT User (IU)" using the PTMF to investigate and model the underlying activities of threat actors responsible for this privacy threat. 

Our research approach begins by developing the PTMF from selected threat \textit{tactics} from the MITRE ATT\&CK framework and \textit{techniques} from the LINDDUN threat model. This makes the proposed PTMF privacy-centered due to the framework techniques selected for the LINDDUN privacy threat model. Based on the PTMF, we conducted a user study to gather the perceptions of privacy and security experts regarding privacy threats in IoT systems. The data from the participant responses are collected, cleaned, and prepared for analysis. These data are then analyzed by mapping the threat actors to the tactics and techniques in the PTMF. Lastly, the results from the analyzed expert response data are used to conduct a quantitative analysis and are presented in a visualized format for clearer representation of our findings. The entire steps involved in this study are presented in \autoref{fig:methology-Process} and each step is explained in details in \autoref{methodology} under \autoref{expert}, \autoref{subsec:systematic} and \autoref{sec:path-analysis}

This work focuses on examining the multifaceted dimensions of threat actors' activities, including threat vectors, threat entry points, threat propagation patterns, potential sensitive data collection, and the broader impact of malicious activities on IoT systems using the PTMF (See \autoref{fig:PTMF}). Using this comprehensive analytical framework, we investigate the prominent threat actors responsible for privacy threats in IoT systems and the techniques these threat actors may employ through a cumulative analysis. Our analysis also examines the activity patterns of threat actors and the propagation of activity within IoT systems through critical path analysis, with the ultimate objective of informing the development of targeted PPTs and enhancing the overall trustworthiness of IoT systems. This proactive approach to understanding the activities of threat actors aims to facilitate more effective threat mitigation strategies and strengthen IoT system privacy structures through enhanced protection for sensitive user and IoT device data in IoT systems.

This work makes significant contributions to the field of IoT privacy preservation by addressing critical gaps in current privacy threat analysis (PTA) methodologies and providing novel insights into the activities of threat actors during privacy threats.

\textbf{Contributions}: The primary contributions of this research are organized as follows:

\begin{itemize}
\item Privacy-Centered Framework Development: We introduce the Privacy Threat Modeling Framework (PTMF). This framework represents a privacy-centric analytical tool specifically designed for the comprehensive assessment of threat actors and their activities in IoT systems (see \autoref{sec:PTMF}).

\item Expert-Informed Empirical Analysis: We conducted a comprehensive user study involving domain experts in privacy and security to evaluate threat actors' activities within IoT systems. This expert consultation provides validated insights into the activity patterns of threat actors. The user study establishes empirical foundations for our analytical framework (see \autoref{expert}).

\item Systematic Threat Actor Characterization: We present an in-depth analysis of prominent threat actors, techniques, collection, and impact in IoT privacy threats in IoT systems. This characterization offers detailed insights into the capabilities and operational patterns of these threat actors, thereby contributing to a deeper understanding of privacy threats in IoT systems (see \autoref{commulative}).

\item Critical Path Pattern Identification: We identify and analyze threat actors' critical path patterns that illustrate the sequential progression of their activities within IoT systems (see \autoref{path}). These patterns provide actionable intelligence for developing targeted countermeasures and inform the strategic deployment of PPT at critical intervention points in the threat lifecycle.
 \end{itemize}

\section{Background}

This section outlines the key concepts and terms used in this analysis by providing a detailed explanation of the MITRE ATT\&CK frameworks and the LINDDUN threat model.

\subsection{MITRE ATT\&CK}
MITRE ATT\&CK is a knowledge-based framework that explains real-world threat actors' tactics and techniques, providing valuable insights for any organization to develop a comprehensive threat model and effective methodologies to mitigate adversarial attacks.
MITRE ATT\&CK describes a threat actor's tactics, techniques, procedures, and common knowledge of an attack, including attributes and patterns gathered from real-world observations \cite{mitre}.
A \textbf{tactic} is a header to a column in a MITRE ATT\&CK framework that explains the intention behind a threat actor's action in a system, while the \textbf{techniques } are how these tactics are executed \cite{rajesh2022analysis}.

\textbf{Why MITRE ATT\&CK?} MITRE ATT\&CK framework provides more granularity and a deep concept of cybersecurity approach as compared to other cybersecurity frameworks such as National Institute of Standards and Technology (NIST) \cite{nist1}, Cyber Kill Chain \cite{yadav2015technical}, Diamond \cite{caltagirone2013diamond}, ISO 27001 \cite{ISO27001}, etc. MITRE ATT\&CK framework's versatility and detailed methodology make it particularly adaptable across various domains, including Enterprise Networks, Mobile Networks, and Industrial Control Systems (ICS), setting it apart from other frameworks that tend to be more specific. MITRE ATT\&CK expresses its tactics similarly to the Cyber Kill Chain, but provides techniques used to achieve these tactics, which are not present in the Cyber Kill Chain \cite{odarchenko2025comparative}. Other frameworks, such as ISO 27001, the NIST Cybersecurity Framework, and the Diamond model, typically focus on attack mitigation rather than the implementation process of these mitigation \cite{odarchenko2025comparative,ferazza2022cyber}.

In this work, we utilize some specific MITRE ATT\&CK tactics in an enterprise matrix applicable to IoT systems, along with the techniques under each tactic derived from the LINDDUN threat catalog \cite{LINDDUNTree}, as shown in \autoref{fig:PTMF}. 
\subsection{LINDDUN PRO threat model}
LINDDUN is a well-known and efficient privacy threat model used for PTA \cite{robles2020linddun}. LINDDUN is an abbreviation that represents Linking (L), Identifying (I), Non-repudiation (Nr), Detecting (D), Data Disclosure (D), Unawareness (U), and Non-compliance (N) \cite{LINDDUNTree}, and all types of privacy threats can be categorized using the techniques of these threat categorization.

The primary reason for obtaining our techniques from the LINDDUN threat tree catalog is that the LINDDUN threat catalog is privacy-focused, unlike techniques in the MITRE ATT\&CK matrix, which is security-focused. This approach enables the selection of appropriate privacy-related techniques for each tactic,  making the entire PTMF privacy-centered.

\section{Related Work}

This section reviews previous studies on Privacy Threat Analysis (PTA) to identify existing gaps in methodologies. We examine previous PTA research based on the LINDDUN privacy framework, the MITRE ATT\&CK security framework, and other established methods.

\subsection{LINDDUN-Based Privacy Threat Analysis (PTA)}
Recent PTA studies have widely adopted the LINDDUN threat model, which is recognized for its maturity and ongoing development \cite{wuyts2014empirical}. Researchers have applied LINDDUN to analyze privacy risks in diverse domains, including Voice-over-IP communications \cite{hofbauer2012conducting}, national identity management systems \cite{nweke2022linddun}, online authentication processes \cite{robles2020linddun}, connected vehicles \cite{chah2022privacy}, Smart Home (SH) \cite{alalade2024privacy}, and personal health information systems \cite{pop00001}. Despite its broad application, LINDDUN-based analyses often lack comprehensive guidance for interdependent privacy scenarios and methods for validating the outcomes of threat modeling, particularly in identifying the activities and intentions of threat actors.

\subsection{MITRE ATT\&CK Framework based Threat Analysis}
The MITRE ATT\&CK framework is a standard for modeling security threats. It has been used to detect multistage cyberattacks \cite{takey2022real}, model threats in cellular networks \cite{rao2023threat}, develop risk-based security frameworks such as SAIF \cite{belfadel2023towards}, and analyze the evolution of IoT malware \cite{chierzi2021evolution}. However, MITRE ATT\&CK is inherently focused on security, and a dedicated privacy-oriented adaptation is needed to analyze the specific tactics and techniques of privacy threat actors.

\subsection{Privacy Threat Analysis (PTA) Using Other Methods}
Other PTA studies have addressed privacy risks in various contexts, such as social network data publishing \cite{ninggal2011privacy}, online advertising \cite{estrada2017online}, smart grid systems \cite{jawurek2011privacy}, and mobile health data collection \cite{iwaya2018mhealth}. While these works identify important privacy threats, they often do not analyze the associated risks to system assets, the threat surface, or how threats propagate through a system.

Based on previous studies on PTA, we observed a significant gap in the literature, as a comprehensive, privacy-focused framework for analyzing the tactics and techniques of threat actors in IoT systems is lacking. To address this, we propose a framework that integrates the LINDDUN privacy threat model with the MITRE ATT\&CK framework to systematically analyze the activities of privacy threat actors in IoT environments.

\section{Privacy Threat Modeling Framework (PTMF)}
\label{sec:PTMF}
\begin{figure*}[ht!]
    \hspace{0.9cm}
        \centering
        \includegraphics[width=19cm,height=9.5cm]{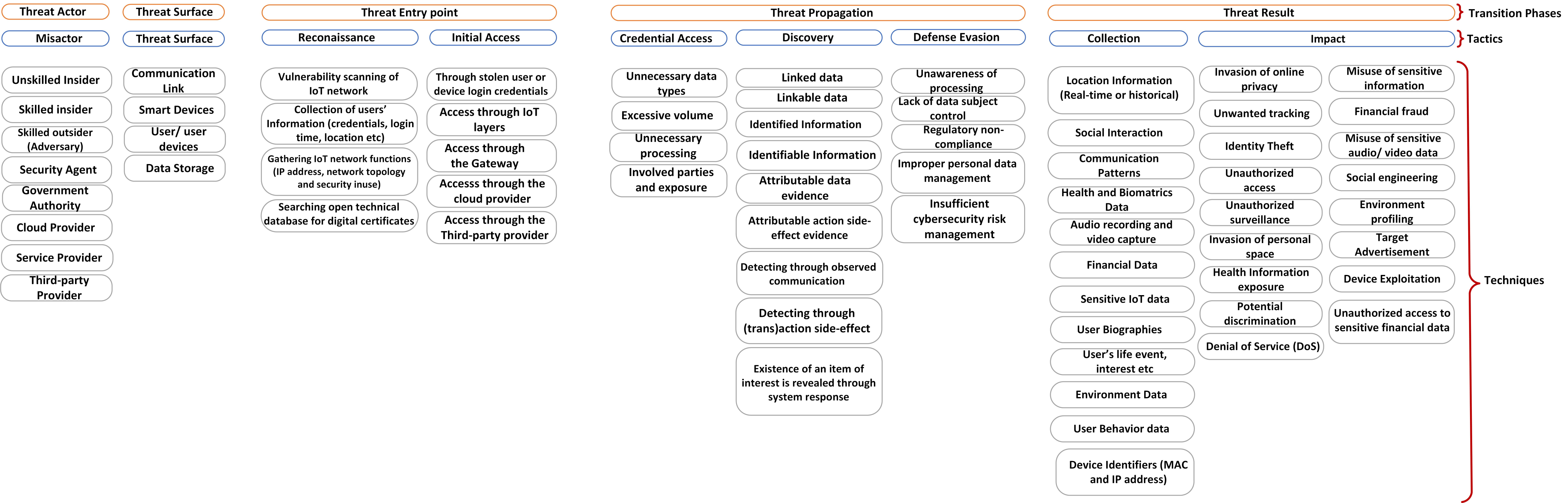}
     \caption{Privacy Threat Modeling Framework (PTMF)}
        \label{fig:PTMF}  
\end{figure*}

The proposed PTMF represents a methodological approach that integrates relevant IoT threat tactics extracted from the MITRE ATT\&CK enterprise matrix. Since techniques under selected MITRE ATT\&CK tactics are security-centered, we employ privacy-specific threat techniques derived from the LINDDUN PRO threat model. This approach makes PTMF a privacy-centric analytical framework specifically designed for assessing the activities of IoT threat actors that could lead to privacy threats. The proposed PTMF analyzes privacy threats through five sequential phases. These phases are: (1) threat actor, (2) threat surface, (3) threat entry point, (4) threat propagation, and (5) threat result as shown in \autoref{fig:PTMF}. The following sections elaborate on the framework's components, detailing the tactics and techniques integrated within each analytical phase.

\subsection{Definitions of Framework Components}

The following are the 5 threat phases and their tactics.
\noindent{\textbf{Threat Misactor}} These are actors involved in the misuse of cases that cause a threat to the privacy of IoT data. In our work, we presented eight misactors in IoT (See \autoref{fig:PTMF}). The skilled insider, skilled outsider (Adversary), and unskilled insider are typically individual actors, while the other misactor encompass organized groups or teams of threat actors, such as malicious organizations  \cite{casey2007threat, arogundade2012towards, walton2006balancing, nurse2014understanding, bugeja2017analysis, meltzer2015nternet, ulltveit2016secure}.
  
\noindent{\textbf{Threat Surface}} These are potential avenues for misuse by malicious actors who may exploit vulnerabilities in these surfaces to perform their malicious acts. In an IoT system, the identified threat surfaces include the communication link, data storage, smart devices, and user/user device. These are referred to as threat vectors \cite{rizvi2020identifying}.

\noindent{ \textbf{Threat Entry point}} Explains ways misactors can access a system based on some of the tactics in the MITRE ATT\&CK framework applicable to our work \cite{mitre}. These tactics are reconnaissance and initial access. The techniques used under each tactic are privacy-specific and obtained from the LINDDUN threat tree \cite{LINDDUNTree}. We considered the following as potential threat entry points for malicious actors to access the IoT system.
    
    \begin{itemize} 
        \item Reconnaissance: The misactor collates useful information that can be used for future operations at this stage. At this stage, threat actors methodically compile actionable knowledge regarding target IoT system architectures, network topologies, and security strength to inform strategic planning for next threat phases \cite{mitre}. 

        \item Initial Access: Following reconnaissance activities, threat actors execute penetration attempts against IoT system infrastructure by leveraging previously gathered intelligence to identify and exploit optimal threat vectors. This phase encompasses the systematic exploitation of vulnerabilities within IoT system components, including the IoT layer, gateways, cloud providers, third-party, and through stolen credentials, to establish an initial foothold within the target IoT system \cite{mitre}.

        \end{itemize}
  After successfully infiltrating the targeted system, misactors tend to engage in activities such as exploration, data gathering, and establishing their presence within the system to ensure continued access. These actions often involve creating a backdoor to facilitate future attacks and the propagation of threats \cite{mitre}.

\noindent{\textbf{Threat Propagation}} 
    At this stage, the misactor has access to the IoT system and is now trying to explore the entire system. Using MITRE ATT\&CK tactics and LINDDUN threat tree techniques, we explain the following threat propagation tactics. \cite{mitre,LINDDUNTree};
    
    \begin{itemize}
     \item Credential Access: Malicious actors employ credential harvesting techniques to obtain legitimate user authentication data, including usernames and passwords, thereby facilitating unauthorized access to and control of systems \cite{mitre}. Within IoT systems, the LINDDUN \cite{LINDDUNTree} privacy threat model identifies several vulnerabilities that can be exploited through vectors that enable credential compromise. Techniques such as collecting unnecessary data types, retaining excessive data volumes, implementing superfluous data processing operations involving multiple parties, and inadvertently exposing sensitive authentication information are some of the processes that enable threat actors to gain unauthorized access to credentials in IoT systems.

    \item Discovery: The misactor aims to gain a deeper understanding of the entire system \cite{mitre}. According to the LINDDUN threat model tree \cite{LINDDUNTree}, privacy-compromising discovery techniques in IoT systems include: device and user profiling through linked datasets, exploitation of linkable data structures, analysis of identified and identifiable information, examination of attributable data patterns, inference from attributable action side-effects, and surveillance of observable communication patterns (see \autoref{fig:PTMF}).
    
     \item Defense Evasion: This category represents process or approach in the IoT system that enables malicious actors to circumvent detection mechanisms and evade defensive countermeasures \cite{mitre}. The LINDDUN framework explains existing privacy-focused evasion approaches that enable misactors to exploit systemic vulnerabilities, including user unawareness regarding data processing activities, inadequate user control mechanisms over personal data, non-compliance with privacy regulations by service providers, the implementation of substandard personal data management practices, and insufficient cybersecurity risk management protocols \cite{LINDDUNTree}.
    \end{itemize} 
  
\noindent{\textbf{Threat Result}}
    Following the threat actors' access and their threat propagation, we considered the effect of their actions on the assets and the entire system. First, we considered possible sensitive data that could be collected, and secondly, the impact these could have on the IoT system.

    \begin{itemize}
        \item Collection: After establishing persistent access and executing threat propagation mechanisms, threat actors initiate the next step, which involves collection tactics. This collection tactic involves the methodical extraction of valuable information assets, with particular emphasis on data categories that directly support privacy invasion campaigns and facilitate advanced persistent threat activities \cite{siboni2018security}. The collection methodology encompasses automated data harvesting techniques, selective targeting of high-value information repositories, and comprehensive system reconnaissance to identify previously undiscovered data sources within the compromised part of the IoT system.
        \item Impact: The consequential outcomes of the threat actor action, which can result in manipulation of data integrity and system configurations, strategic interruption of critical business processes and operational workflows, and comprehensive destruction of system infrastructure and information assets \cite{rajesh2022analysis}. The magnitude and persistence of these impacts are determined by the threat actor's intentions, available resources, and the defensive posture of the targeted IoT system \cite{mitra2021impact}.

         The phases, the tactics, and possible techniques the threat actors could use for a successful IU privacy threat in IoT systems are analyzed through the PTMF in \autoref{fig:PTMF}.
    \end{itemize}

\subsection{Application of PTMF}
The proposed PTMF can be utilized for various privacy preservation purposes; some examples of PTMF applications are as follows:
\begin{itemize}

    \item Analysis of Threat actors' activities in IoT systems: A dataset related to privacy breaches can be analyzed to reveal how threat actors' activities can be identified during privacy threats, utilizing the PTMF tactics and techniques. However, these datasets are not publicly available for analysis, due to sensitivity of these data \cite{abraham2025promoting}. However, a user study could be conducted to gather data from privacy and security experts about their opinions on privacy threats in IoT systems. The data in both cases could be used to map threat actors to techniques under each tactic in \autoref{fig:PTMF}. This will illustrate the variations in the pattern of threat actors' activities associated with each privacy threat affecting IoT systems. In this work, we conducted an expert-driven user study based on PTMF, and the methodology of the process is presented in detail in \autoref{methodology}.
    
    \item Risk Assessment of IoT Threat Surface: The PTMF framework can be utilized for threat surface risk assessment in large IoT systems, such as the Internet of Industry (IIoT). The IoT administrator managing this IoT system can be provided with a questionnaire template on the current privacy measures in place for their IoT system to mitigate threat actors' techniques during reconnaissance, initial access, credential access, discovery, and defense evasion, as shown in \autoref{fig:PTMF}. Based on the IoT administrator's response to these questions, the assessment of the privacy risk of the IoT system can be calculated and presented in the form of a heatmap. This approach will provide the IoT administrator with insight into which threat surfaces are more vulnerable and require effective implementation of privacy measures. 

    \item Privacy Threat Analysis System (PTAS): The user study above could be expanded to collect possible privacy mitigation methods for each technique under each tactic. The mapping of threat actors' techniques and possible mitigations per technique to threat results (collections and impact) could be used to develop an effective privacy threat analysis system (PTAS). PTAS could be used by IoT users to analyze the level of privacy threat risk associated with their IoT system by revealing the possible sensitive data that could be collected (assets) and the potential impact if this sensitive data is misused. This will enable IoT user to be proactive in mitigating the type of privacy threats their IoT system are prone to.
\end{itemize}
With various ways PTMF can be utilized, this work presents an analysis of the privacy threats in IoT systems, examining the threat actors responsible for these privacy threats and their activities within IoT systems. This application of the PTMF is explained in detail in the following section.
\section{Threat actors' activities in IoT systems}
\label{methodology}
 
In this section, we explain one of the applications of PTMF in \autoref{fig:PTMF} by analyzing the activities of threat actors in IoT systems based on the tactics and techniques employed by these threat actors to carry out privacy threats in IoT systems. We will be analyzing twelve (12) privacy threats associated with IoT systems as gathered from the literature. This section of our work is dedicated to the Identification of IoT user (IU) privacy threats in IoT systems. Moreover, we present the results of the remaining 11 privacy threats in Appendix \ref{appendix:A}. 

\textbf{Identification of IoT User (IU)} is one of the privacy threats that pose risks to privacy in IoT systems. It occurs when threat actors are able to link sensitive information with specific individuals. These connections can significantly compromise user privacy by not only revealing sensitive user data but also connecting it with other personally identifiable information (PII). This vulnerability primarily results from inadequate data minimization practices, which increase the likelihood of unintentionally disclosing excessive amounts of information \cite{seliem2018towards, ziegeldorf2014privacy, ogonji2020survey, shaikh2019internet, alhalafi2019privacy, al2016overview, jain2020privacy}.

\begin{figure}[ht]
        \centering
        \includegraphics[width=9cm,height=6cm]{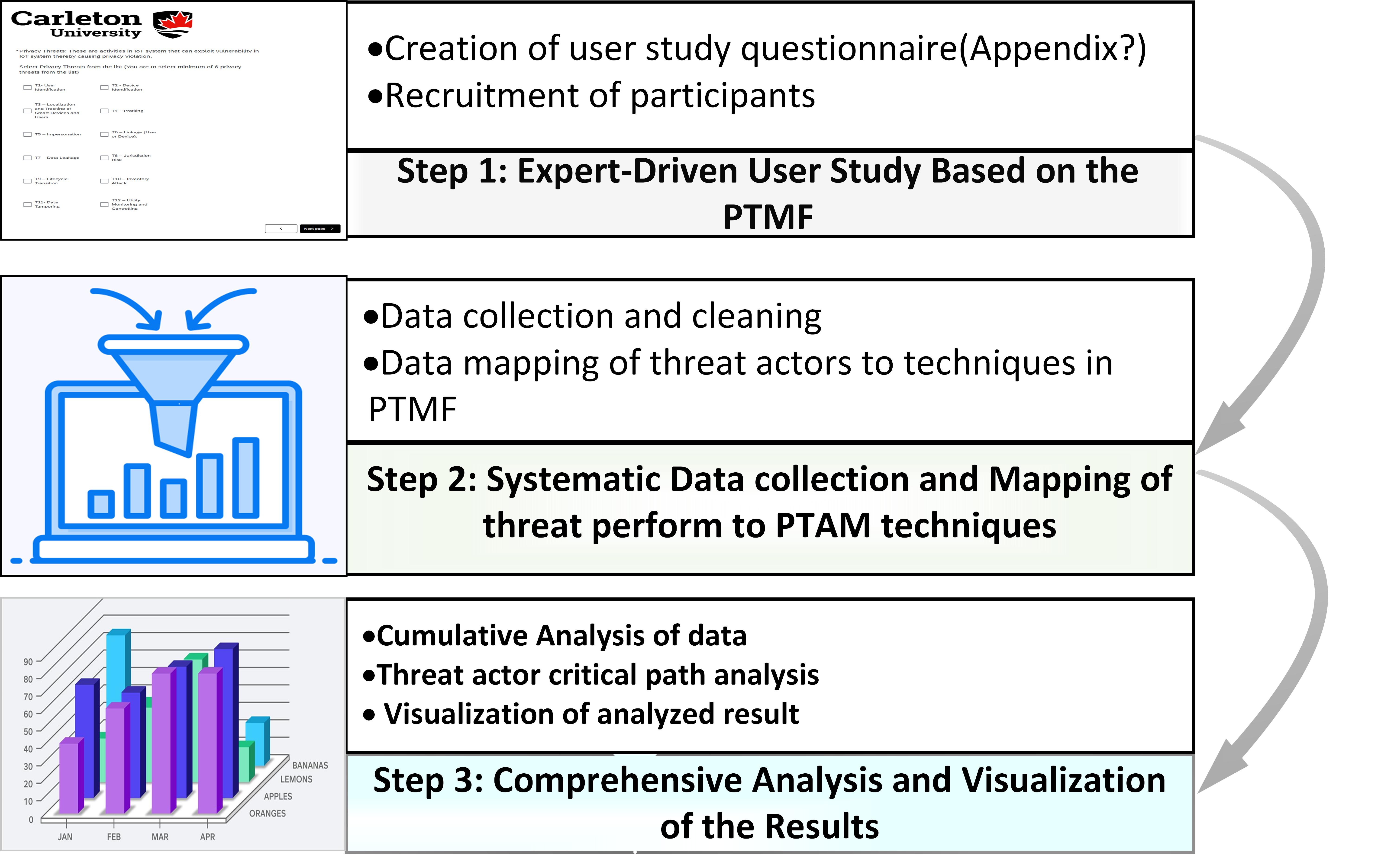}
        \caption{Methodological Workflow for the user study process and evaluation of the activities of  threat actors  responsible for privacy threats in IoT systems}
        \label{fig:methology-Process}
  \end{figure}

For an effective explanation of the utilization of PTMF in privacy threat analysis, we explain the process involved in our work as follows: 1) we conduct an expert-driven user study based on PTMF in \autoref{subsec:expert}, 2) we perform systematic data collection and mapping of threat actors to PTMF techniques in \autoref{subsec:systematic}, and 3) present the threat actor critical path analysis in \autoref{sec:path-analysis}. These processes summarize the methodology of our work, as shown in \autoref{fig:methology-Process}.
\\

\subsection{Expert-Driven User Study Based on the PTMF}
\label{expert}
\subsubsection{The User Study Design and Testing Process}
The user study in this work was conducted to address the less focused factor involved in the occurrence of privacy threats in IoT systems, which is the threat actors. To obtain detailed information about these threat actors, how they could gain entry points into the IoT systems, and their propagation within the IoT systems, we conducted an anonymous online user study approved by \href{https://carleton.ca/researchethics/cureb-b/}{CUREB} with Ethics Clearance ID \# \textbf{122487}.
 The user study is grounded in the PTMF presented in \autoref{fig:PTMF}, which aims to capture expert insights into privacy threats in IoT systems. A structured questionnaire was developed using Qualtrics Experience Management (XM), an online survey platform, to systematically gather expert responses \cite{qualtrics}.

\begin{figure}[ht]
        \centering
        \includegraphics[width=8.5cm,height=5cm]{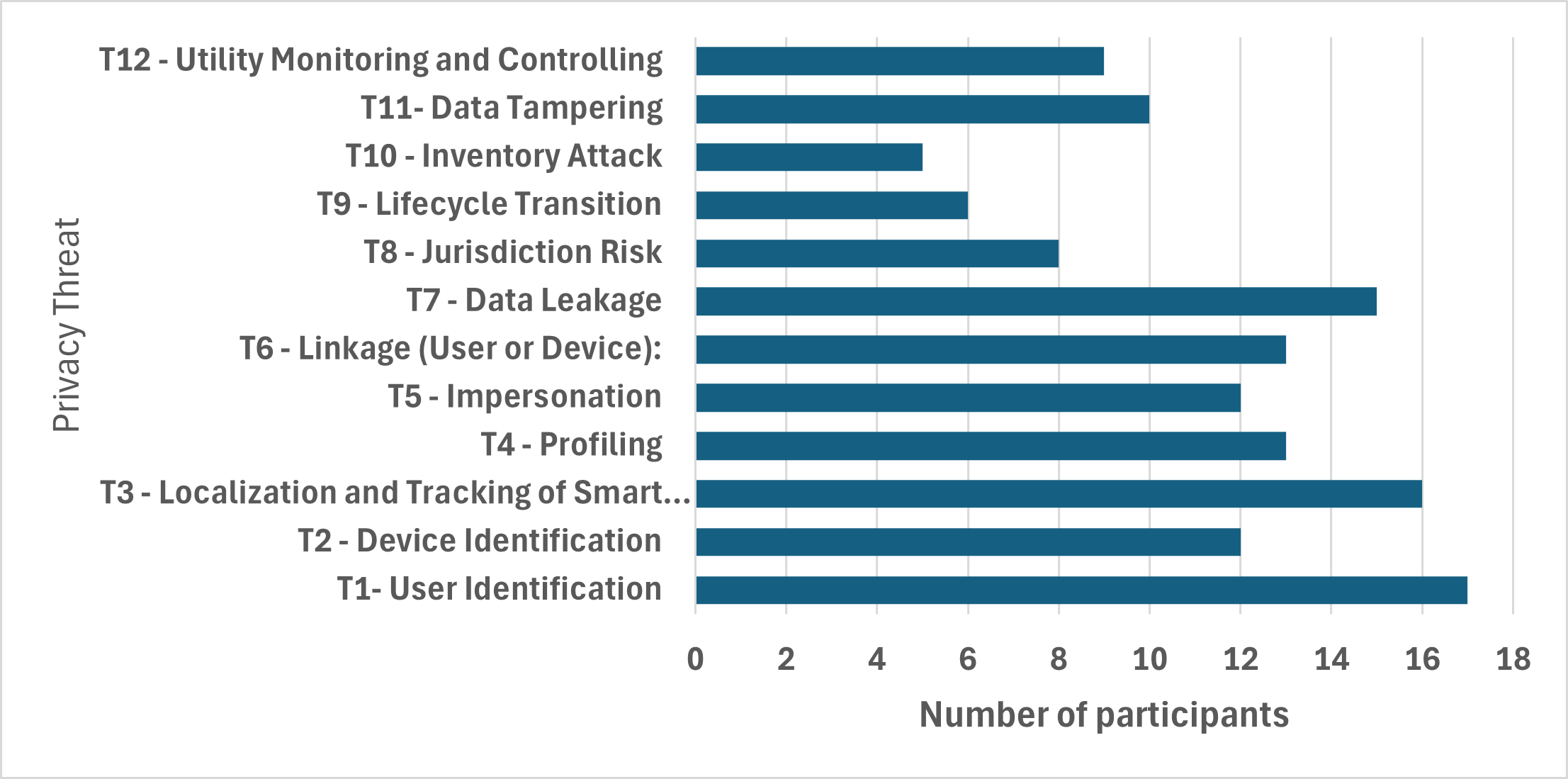}
        \caption{The variation in the number of participants for each privacy threat}
        \label{fig:threat_selection}
 \end{figure}

The questionnaire for the user study was segmented into 13 parts, with the first part focusing on participant demographics. The remaining 12 parts are allocated to each privacy threat covered in our study. The total number of questions for the user study is 931, with 77 questions allocated for each of the 12 privacy threats and 7 questions under demographics. We present the questionnaire for review by an expert in quality assurance and colleagues for user experience, as well as for any inconsistencies in the questionnaire. Due to the volume of questions, we conducted a study for a specified duration to complete the demographic part and one of the privacy threat parts. It took an average of 2 minutes to complete the demographics and 13 minutes for the selected privacy threat, making a total of 15 minutes. To make the user study flexible enough for our participants, we allowed each participant to choose a minimum of 6 out of the 12 privacy threats. The participant selection distribution for each privacy threat is presented in \autoref{fig:threat_selection}. Our analysis for each privacy threat is based on the participants' opinions on each threat, with the number of participants per threat varying as shown in \autoref{fig:threat_selection}.

\subsubsection{Recruitment and Participant-Provided Data Integrity}
We initiated the recruitment process by inviting experts in privacy and security from both academia and industry. Each participant was presented with a predefined list of questions to gauge their insight into privacy threats, as outlined in the PTMF, and asked to select the most significant techniques under each tactic based on their experience. Subsequently, participants were guided to assess each selected threat across the various phases defined in the PTMF, namely: threat actors, threat vectors, threat entry points, threat propagation, and threat collection. For each phase, participants were prompted to identify potential techniques that the selected threat actors could realistically employ within an IoT system. This expert-driven input formed the foundational dataset for subsequent analysis and mapping of threat actors to techniques.

To validate the accuracy and usefulness of the information provided by participants, we included a qualifying question at the end of the survey for participants to summarize what they learned and offer suggestions about the user study. This gives us insight that the participant takes the time to thoroughly review all the questions and answer them based on their level of expertise in privacy and security, rather than simply answering the questions arbitrarily.

\subsubsection{Participant Demographics}
This section provides an overview of the demographics of all participants, including their country of residence, education, career sector, years of experience, and level of expertise in privacy and security. These demographics show that we have fewer privacy experts compared to security experts; however, based on the minimum years of experience, we were able to get a response from an experienced expert in this field. 


\begin{itemize}
    \item {Country of Residence Distribution}: The majority of the participants are from North America, with 65\% from Canada, 30\% from the  United States, and 5\% from Uni ted Kingdom 

\item{Level of Education Distribution}: Majority of the participants (95\% ) have postgraduate degrees, such as a Master's and a Doctorate's degree.

 \item{Year of Experience Distribution}: We have a large percentage of participants within the 3 and 6 year experience range (70\%), which satisfies the level of privacy and security experience required for this user study.
\item {Organization Sector Distribution}: The recruitment was focused on participants from either academia or industry to see how privacy is being embraced in these sectors.

\end{itemize}

\subsubsection{User study Validation}
\begin{itemize}
    \item Sampling size validation: We encountered some challenges during the recruitment process, as there were limited available privacy and security experts willing to participate in the user study, partly due to the long duration of the study, which contributed to the participants' unwillingness to participate. However, to justify the effectiveness of the number of participants involved in this study, we employ a priori power analysis for sample size justification and the effective size of the expert participant opinion, using sensitivity power analysis, as described in \cite{lakens2022sample}. We employ G*Power \cite{gpower} to conduct the following: 1) a sample size test and power analysis for a one-sample t-test to justify the effectiveness of 20 experts recruited for this study. We use the t-test because we are comparing expert opinion to the known possible tactics and techniques presented in PTMF \autoref{fig:PTMF}. As mentioned by Kang et al. \cite{kang2021sample}, sample size and power analysis are based on effective size (d), significant level $(\alpha)$,  power $(1-\beta)$, and statistical analysis type. We conduct a sample size with the test family = t-test, statistical test = Means: Difference from constant (one sample case), power analysis = A priori. The following input parameters where selected; one-tailed, $\alpha$ = 0.05, $d$ = 0.8 which is large effective size and the power $1-\beta$ = 0.95. Calculating the sample size based on these input parameters gives a sample size of 19, which is lower than the sample size used in our user study. 2) We calculate the sensitivity power, with the following input parameters: one-tailed, $\alpha$ = 0.05, sample size = 20, power $1-\beta$ = 0.9, we got a large Cohen's effective size $(d)$ of 0.7636. The high expectation of effect in experts' opinions on privacy threat analysis in IoT systems justifies our sample of 20 experts as suitable for this study, despite the challenges faced during recruitment in this expert-driven research \cite{lakens2022improving, lakens2022sample}.
   
   \item  Participant Privacy and Security Expert Level: We introduce a percentage sliding scale to determine the privacy and security skill level of the user study participants. The skill level of all participants was assessed before they could proceed with the user study, and we obtained a satisfactory level of skill from the participants. On average, participants demonstrated security skill levels of 86.6\% and privacy skill levels of 69.5\%.
   
\end{itemize}

\label{subsec:expert}


\subsection{Systematic Data Collection and Mapping of
threat actor to PTMF techniques}
\label{subsec:systematic}
 
Following the collection of the data from the user study, we conducted a thorough data cleaning process on the information collected from user study participants to ensure reliability and remove inconsistencies. Once the data were prepared, we conducted a focused mapping procedure aligned with the PTMF, as shown in \autoref{fig:PTMF}. This involved linking selected threat actors to relevant threat vectors and specific techniques under entry point tactics (reconnaissance and initial access) and threat propagation tactics (credential access, discovery, and defense evasion).

Additionally, each threat actor was mapped to the types of sensitive data they target within IoT systems and the associated impact of potential compromise. We then identified the most persistent threat actors and their corresponding techniques, data targets, and impact levels based on the frequency of occurrence in the collected dataset. This step enabled a focused understanding of the most critical threat actors, threat vectors, and techniques used for the IU privacy threat in IoT systems. These threat actors' critical paths were visualized for a clearer presentation of the results.




\section{Comprehensive Analysis and Visualization of the Results}
\label{sec:path-analysis}

The quantitative analysis of this work is carried out by conducting 1) a cumulative impact analysis of threat actors and corresponding techniques in the PTMF, and 2) an analysis of threat actors' activities and corresponding critical paths.

\subsection{Cumulative Impact Analysis of Threat Actors and Corresponding Techniques in PTMF}
\label{commulative}
\begin{figure*}[ht!]
        \includegraphics[width=19cm,height=8.5cm]{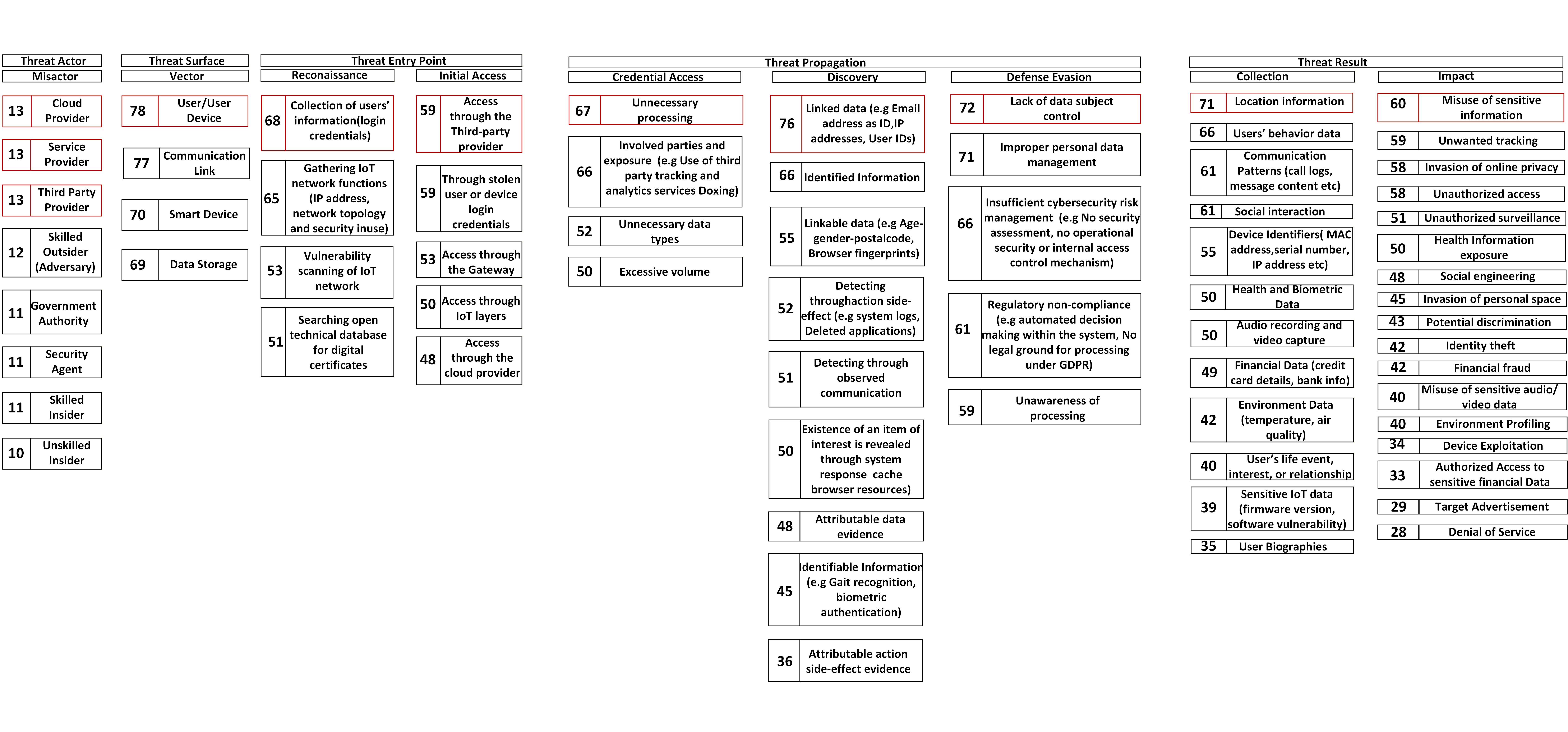}

        \caption{Highlight in red box the progression of threat actors' activities from initial attack vectors through system infiltration to the compromise and misuse of sensitive information.}
        \label{fig:Mapping}
  \end{figure*}

The cumulative approach in this work quantifies the significance of each threat actor and their techniques by summing the total number of responses from user study participants. The frequency of each response is shown in \autoref{fig:Mapping} in descending order.
  
To further map the top threat actors to the top techniques under each tactic, we indicate these top threat actors and techniques in red, highlighting the top threat actors and possible techniques they could employ to identify users of an IoT system (See \autoref{fig:Mapping}). 

\textbf{Our Observation}: Based on the red-highlighted boxes in the PTMF diagram  in \autoref{fig:Mapping}, the following observations can be made regarding the primary threat actors and their activities:

The analysis reveals that cloud providers, service providers, and third-party providers constitute the predominant threat actors within the IoT ecosystem. These entities primarily exploit vulnerabilities through user/user device interfaces to gain unauthorized access to user information. The threat propagation analysis reveals that these actors utilize reconnaissance techniques as an initial entry point, collecting users' data, which is subsequently processed and aggregated without explicit user consent or control.

Furthermore, the threat characterization indicates that these activities, while privacy-invasive, may not be malicious in intent, as the threat actors (cloud provider, service provider, and third-party provider) in this privacy threat may not be deliberately seeking to harm IoT users upon identification. For example, the data collected by the three prominent threat actors may subsequently be used for location tracking, behavioral profiling, and targeted advertising purposes. 


\subsection{Analysis of Threat Actors' activities and corresponding Critical Paths}
\label{path}

\begin{figure*}[ht!]
        \centering
        \includegraphics[width=19cm,height=9cm]{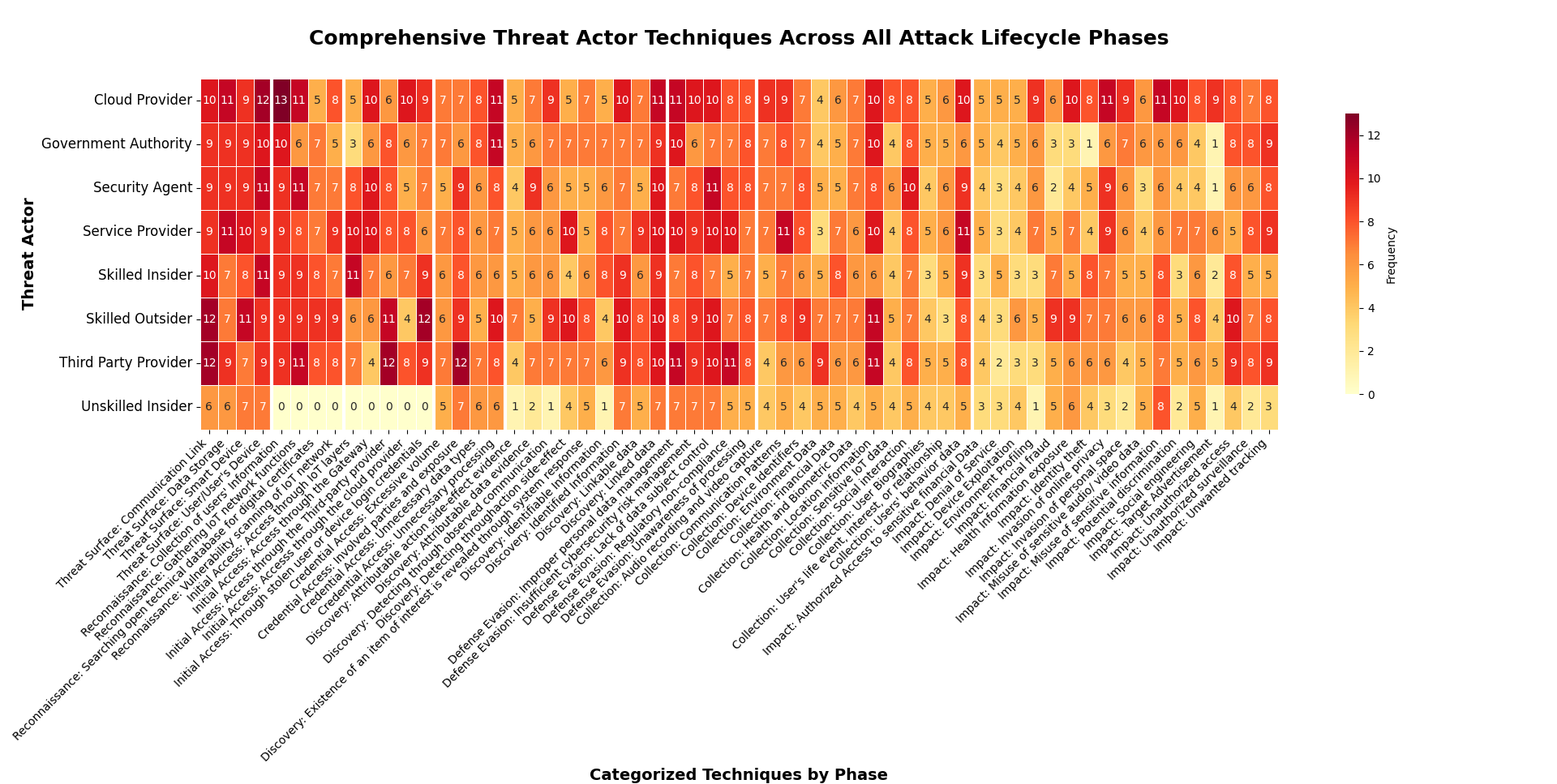}
        \caption{Frequency of privacy threat techniques used by different threat actors. The heatmap illustrates various techniques (x-axis), categorized by threat tactics, versus the type of threat actor (y-axis). The color intensity of each cell, along with its numerical value, represents the frequency with which a specific actor is likely to employ a given technique.}
        \label{fig:Heatmap1}
 \end{figure*}

\begin{figure*}[ht!]
        \centering     \includegraphics[width=18cm,height=9cm]{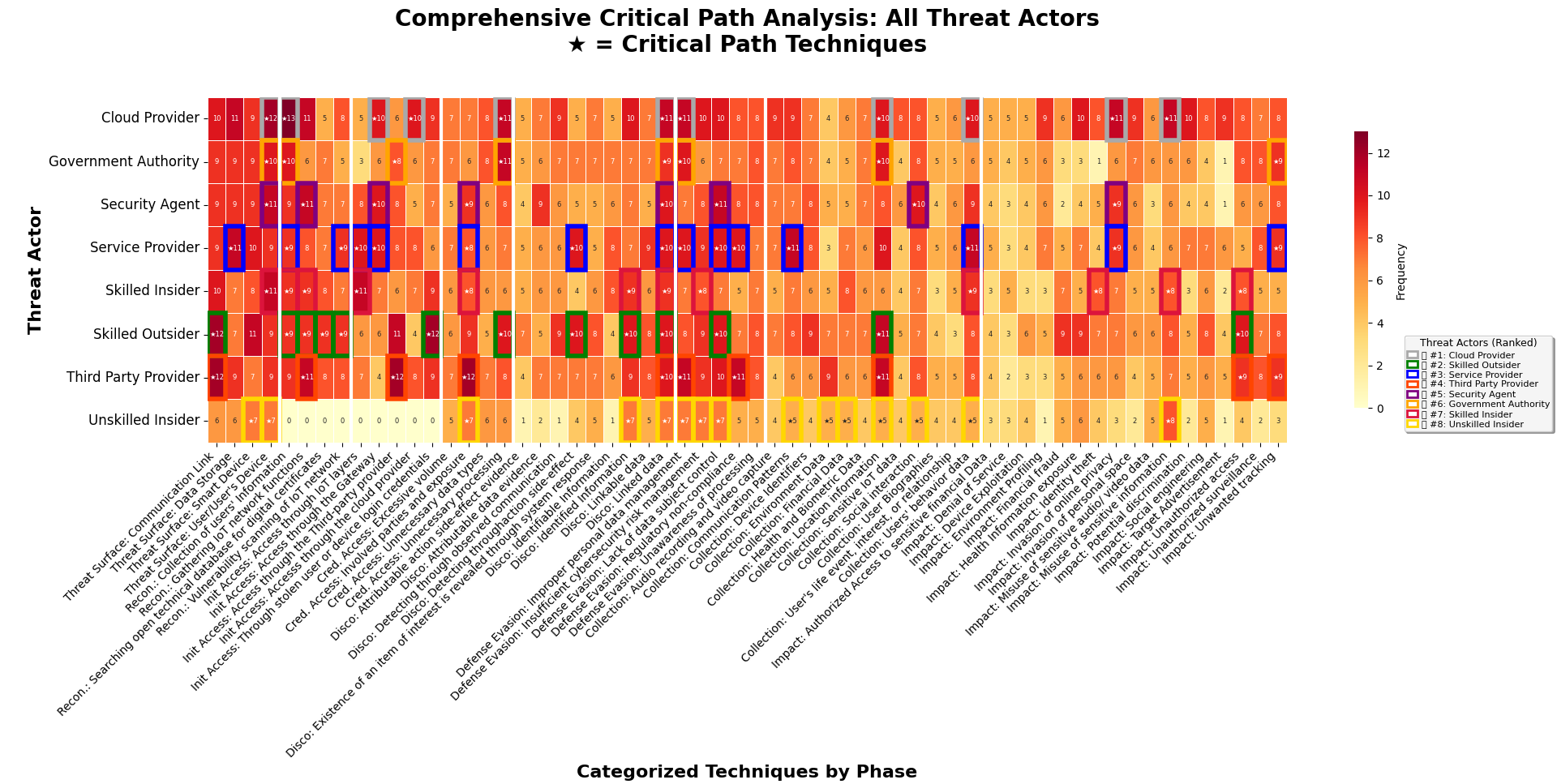}
        \caption{Frequency of privacy threat techniques used by different threat actors. The heatmap illustrates techniques (x-axis) categorized by threat tactics versus the type of threat actor (y-axis). The color scale indicates the frequency of technique usage. Starred ($\star$) cells denote techniques that are part of a critical path.}
        \label{fig:Heatmap2}
 \end{figure*}

\begin{figure*}[ht!]
\includegraphics[width=18.5cm,height=11cm]{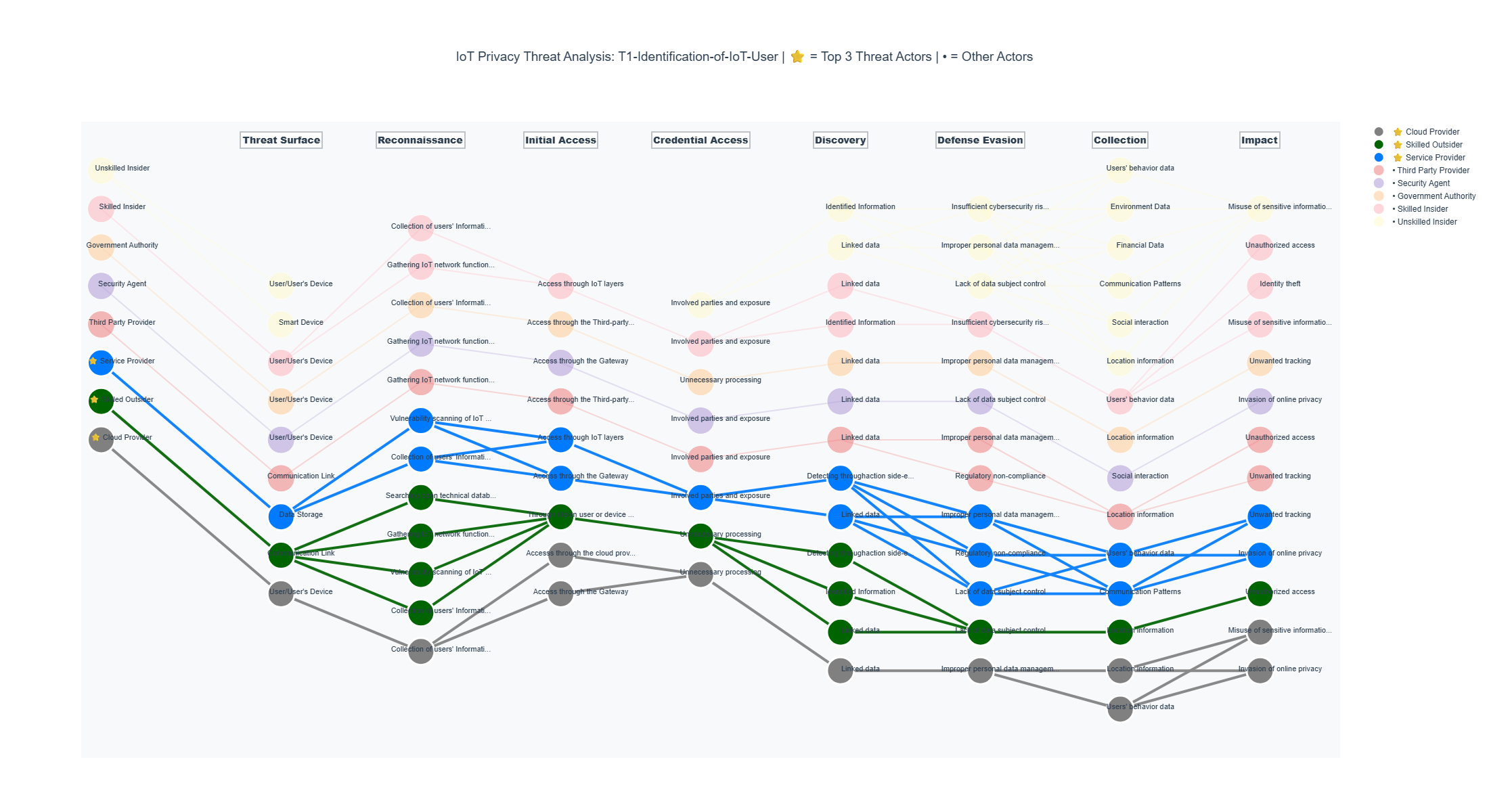}
\caption{A network graph showing the Threat Actors techniques across all Threat Tactics}
\label{fig:Net-graph1}
\end{figure*}

To gain a deeper understanding of threat actor activity during IU privacy threats, we examine the leading threat actors and their associated techniques by mapping out the critical attack path. This critical path is illustrated in the heatmap presented in \autoref{fig:Heatmap1}, where we highlight the most significant cells that correspond to the key techniques used by each threat actor. The following is the process used in identifying the critical path in our work.
\begin{itemize}
    \item Critical Path Identification: 
     The dataset is segmented by threat tactic. Within each tactic, the technique(s) with the highest frequency are identified as critical path components. However, multiple techniques sharing the highest frequency under a tactic are all included. The critical path identification process is a tactic-wise mapping of each threat actor’s most significant threat techniques.
     \item Critical Path Highlighting: We assigned distinct border colors to all the threat actors for their critical path techniques (see \autoref{fig:Heatmap2}). For example, the top three threat actors' colors are indicated as follows: \colorbox{Gray}{Gray} for the highest-ranked actor ( cloud provider), \colorbox{Green}{Green} for the second-ranked   ( skilled outsider), and \colorbox{Blue}{Blue} for the third-ranked actor ( service provider). Other color indications are \colorbox{Tan}{orange} denotes government authorities, \colorbox{lavender}{purple} indicates security agents, \colorbox{verylightsalmonpink}{pink} represents skilled insiders, \colorbox{salmonpink}{peach} indicates third-party providers, and \colorbox{yellow}{yellow} represents unskilled insiders.
\item Critical path cells: The critical path cells are annotated with a star ($\star$) and the frequency value for immediate identification (see \autoref{fig:Heatmap2}).
\item Critical path analysis: We further express all the threat actors and corresponding critical path in a network graph for optimized readability in \autoref{fig:Net-graph1}. We highlight the top 3 threat actors to increase their visibility and critical techniques during the critical path analysis visualization in \autoref{fig:Net-graph1}.

\end{itemize}

\textbf{Our Observation: } Based on the critical path analysis visualization in \autoref{fig:Heatmap1} and \autoref{fig:Heatmap2}, several key observations emerge from the highlighted nodes representing the most critical threat pathways. The analysis reveals that threat actor selection correlates directly with technique frequency, where the most frequently employed techniques across each tactic establish the primary critical path nodes.



An examination of the critical path associated with these three primary threat actors reveals convergent operational patterns, particularly in their shared use of a common vulnerable threat surface, techniques, data types collected, and their impact. However, the skilled outsider demonstrates a distinctive deviation from this pattern, as evidenced by the unique pathway terminating in "unauthorized access", a characteristic that differentiates this threat actor's activity and intent from those of the other two primary actors. This unauthorized access objective represents a fundamentally different threat model compared to commutative analysis in \autoref{fig:Mapping}, despite both analysis approaches having cloud and service providers as part of the top three privacy threat actors responsible for IU privacy threats. Service provider threat actors aim to conduct "unwanted tracking" of user locations, which could be utilized for marketing purposes by directing advertisements to users based on their location. However, cloud providers can "misuse users' sensitive information" beyond the initial intended use of the collected data. This act usually occurs when sharing data with other providers, such as service or third-party providers. Furthermore, the invasion of online privacy is common to both cloud and service providers, who monitor user activities through their data, and based on this data, they could suggest new services in the form of advertisements.

In summary, this critical-path PTA approach focuses on analyzing threat actor activities through our novel PTMF. We carry out our empirical evaluation of the PTMF by analyzing the data collected from our user study participants. The result from the analyzed data shows the critical paths that threat actors could use to identify users in an IoT system, and the top three (3) prominent threat actors. Furthermore, the PTMF can be applied to evaluate other privacy threats that could impact IoT systems. To demonstrate the broader applicability of PTMF, we analyzed eleven additional IoT privacy threats identified from the literature using data collected from user study participants. The comprehensive results of this extended analysis are presented in Appendix  A, and a table summarizing the top 3 threat actors and techniques for each privacy threat is included in Appendix B. Also, the interactive visualization of the network graph for the twelve (12) privacy threats covered in our work can be accessed from this \href{https://iot-threat-dashboard.onrender.com/}{link}.

\section{Discussion}

The analysis in this work reveals prominent privacy threat actors responsible for identifying IoT user identities in IoT systems, as identified through the analysis of data collected from user study participant responses and the visual presentation of results based on static data frequency. 

The cumulative analysis shows the mapping of prominent threat actors and techniques based on the highest frequency of responses from the participants. However, the critical path-based selection identifies prominent threat actors based on the frequency of techniques associated with them. The cumulative approach presents its results based on the threat actor cumulative frequency value, and the top threats are cloud provider, service provider, and third-party. In contrast, our critical path analysis approach identifies the top threats as the cloud provider, the service provider, and the skilled outsider. 

The most common techniques, as gathered from the user study participant responses, reveal a critical pattern that threat actors could use to expose user identities in an IoT system. The comparison results show that cloud providers are the prominent threat actor in both analysis approaches, as shown in \autoref{fig:Mapping} and \autoref{fig:Net-graph1}. Our cumulative analysis reveals that cloud providers are the predominant threat actors, followed by service providers, and then third-party providers. However, the critical-path analysis is slightly different, with the prominent threat actor being the cloud provider, followed by the skilled outsider, and then the service provider. This analysis shows the threat actors that can be responsible for the IU privacy threat in an IoT system; therefore, the privacy mitigation process should prioritize cloud providers, service providers, skilled outsiders, and third parties, as revealed in our analysis.  \\

\textbf{Limitation and Future Direction}:
The privacy threats analyzed using the PTMF may exclude certain privacy threats not covered in the literature, particularly those that are new or emerging. Furthermore, the user study should also include other IoT stakeholders, such as IoT users, IoT administrators, IoT installer, and their experiences with privacy threats.
In addition, the future direction of this work can be extended by applying PTMF within automated security tools, which presents an opportunity to enhance real-time threat detection and response. This includes developing PTMF-based modules for privacy assessment platforms and collaborating with industry partners to integrate these modules into the risk management processes of IoT manufacturers and providers. 

\section{Conclusion}
\label{conclusion}
This work presents a systematic methodology for assessing and analyzing privacy threats in IoT systems, with a focus on the development and application of the Privacy Threat Modeling Framework (PTMF). Leveraging selected tactics from the MITRE ATT\&CK framework and techniques from the LINDDUN threat model, we developed a privacy-centric PTMF that provides a comprehensive structure for identifying, mapping, and analyzing the activities of threat actors relevant to IoT privacy threats. 

Through an expert-driven user study and a systematic mapping of threats and techniques, we identified the most persistent threat actors, prioritized key techniques, and critical paths that facilitate IU privacy threats and other privacy threats in IoT systems. Comprehensive quantitative and visual analysis further elucidates the relationships among threat actors, tactics, and techniques, providing actionable insights for researchers and practitioners into areas of privacy in IoT systems that require further research. This analysis reveals the cloud provider as the most prominent threat actor that could pose a risk to IU's privacy in IoT systems. Service providers, skilled outsiders, and third parties are other prominent threat actors that could be responsible for IU privacy in IoT systems.

Overall, the proposed methodology not only advances understanding of privacy threat dynamics through analysis of threat actors' activities in IoT systems but also provides a flexible foundation for future threat modeling and the development of targeted privacy-preserving controls.

\bibliographystyle{IEEEtran} 
\bibliography{Ref}

\section*{Appendix A}	
\label{appendix:A}
\subsection{T2-Identification of IoT device}
\label{T2}
It occurs when threat actors can associate identifiable attributes of a device, such as its IP address, firmware version, serial number, etc \cite{seliem2018towards, ziegeldorf2014privacy, ogonji2020survey, shaikh2019internet, alhalafi2019privacy, al2016overview, jain2020privacy}.

 \begin{figure*}
        \centering
    \includegraphics[width=19cm,height=10cm]{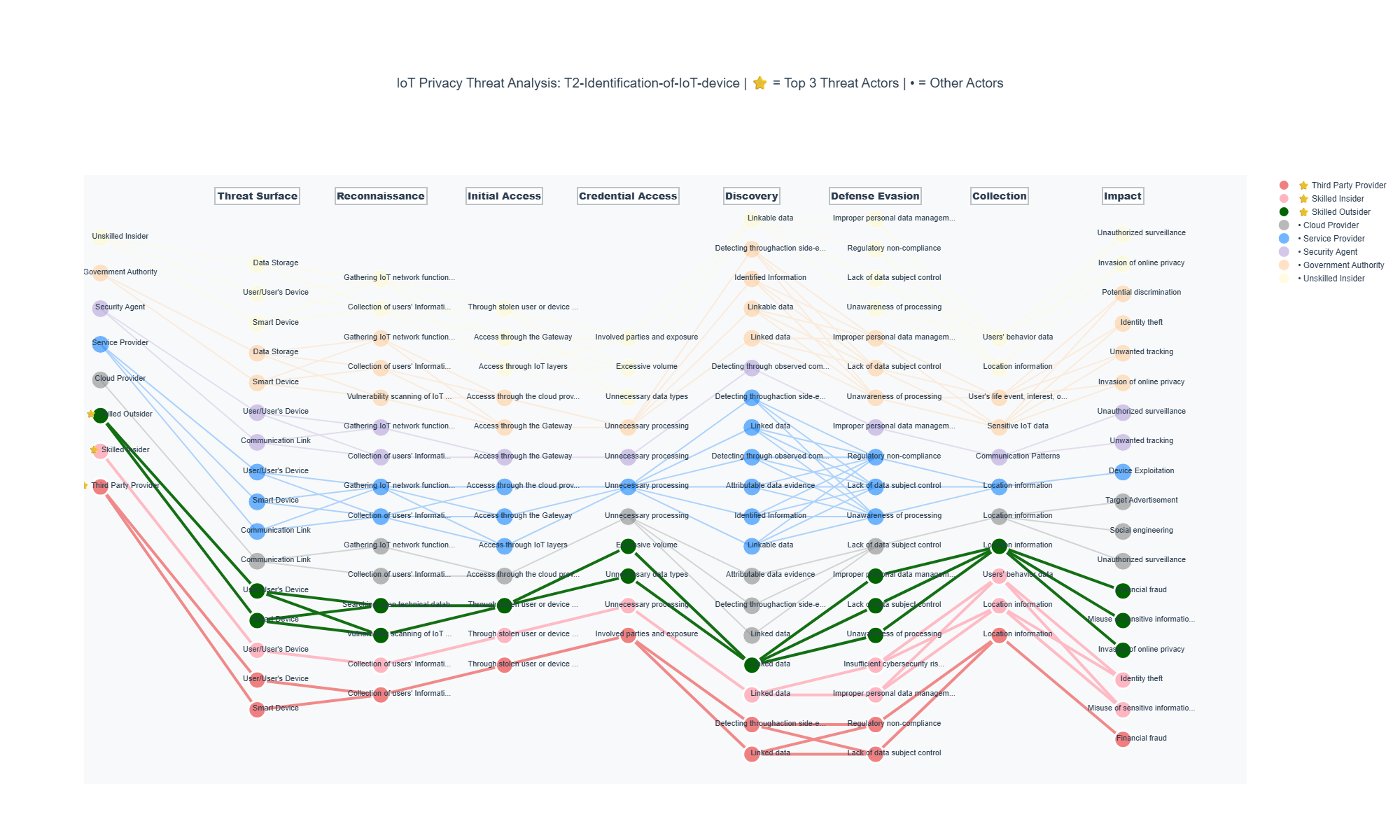}
        \caption{T2-Device Identification network graph showing the Threat Actors techniques across all Threat Tactics}
        \label{fig:T2Net-graph1}
 \end{figure*}

\textbf{Our observation}: Focusing on the top 3 threat actors in \autoref{fig:T2Net-graph1}, the intentions of cloud and service providers are not malicious, as their aim is to invade IoT users' online privacy through the IoT device for purpose of business development through customer-focused optimization processing of demography data and interest data  \cite{pinto2024systematic} \cite{menard2020analyzing}. The cloud provider intends to misuse sensitive information of IoT devices and users beyond the initial aim of collecting the information. Likewise, the service provider tends to track the location of the IoT device or the user in the case of a wearable IoT device. These acts by cloud and service providers could lead to a lack of trust among the IoT user community. On the other hand, skilled outside intentions are usually malicious compared to the other top two threat actors, as skilled outsiders aim to gain unauthorized access to the IoT system to exploit more vulnerabilities within it.

\subsection{T3-Localization and tracking of smart devices and users}
\label{T3}
It is the continuous monitoring of a user or device's location over time and gathering information about their movements \cite{seliem2018towards, ziegeldorf2014privacy, ogonji2020survey, shaikh2019internet, alhalafi2019privacy, al2016overview, jain2020privacy}.

 \begin{figure*}
        \centering
    \includegraphics[width=19cm,height=11cm]{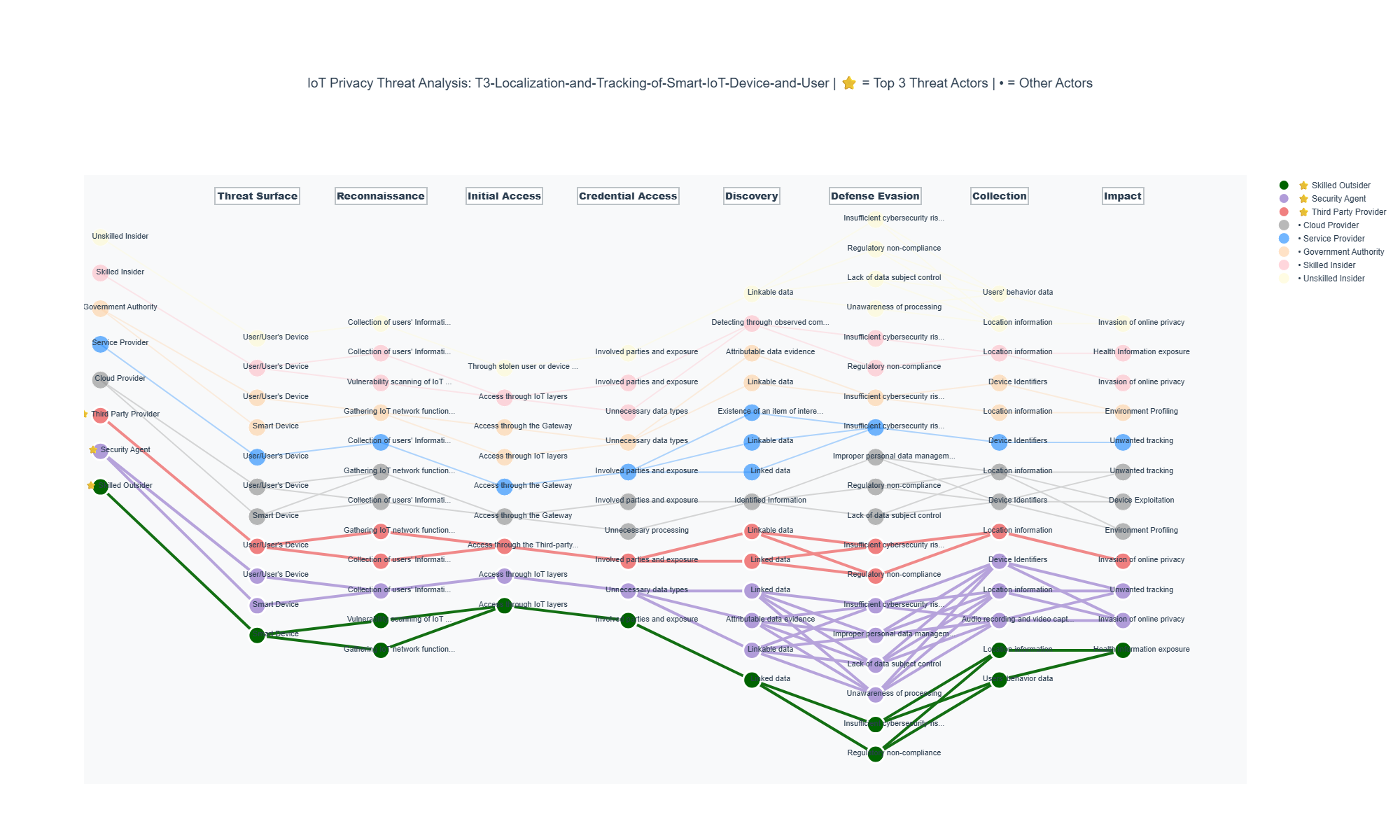}
        \caption{T3-Localization and tracking of smart devices and users network graph showing the Threat Actors techniques across all Threat Tactics}
        \label{fig:T3Net-graph1}
 \end{figure*}

\textbf{Our observation}: The intention of the top 3 threat actors that are mostly responsible for the localization and tracking of smart devices and IoT users can be malicious or non-malicious.For example, the skilled outsider typically intends to collect sensitive user information, such as health information. However, the security agent's intention is to identify the device and the user's location for investigation purposes, which could be considered an invasion of the user's online privacy and unwanted tracking. Nevertheless, this act is necessary due to the target IoT device and the user involved.

\subsection{T4-Profiling}
\label{T4}
It leverages compiled and correlated data about users' activities that have been analyzed to infer their behavior and personal interests \cite{seliem2018towards, ziegeldorf2014privacy, ogonji2020survey, shaikh2019internet, alhalafi2019privacy, al2016overview, jain2020privacy}.

  \begin{figure*}
        \centering
    \includegraphics[width=19cm,height=11cm]{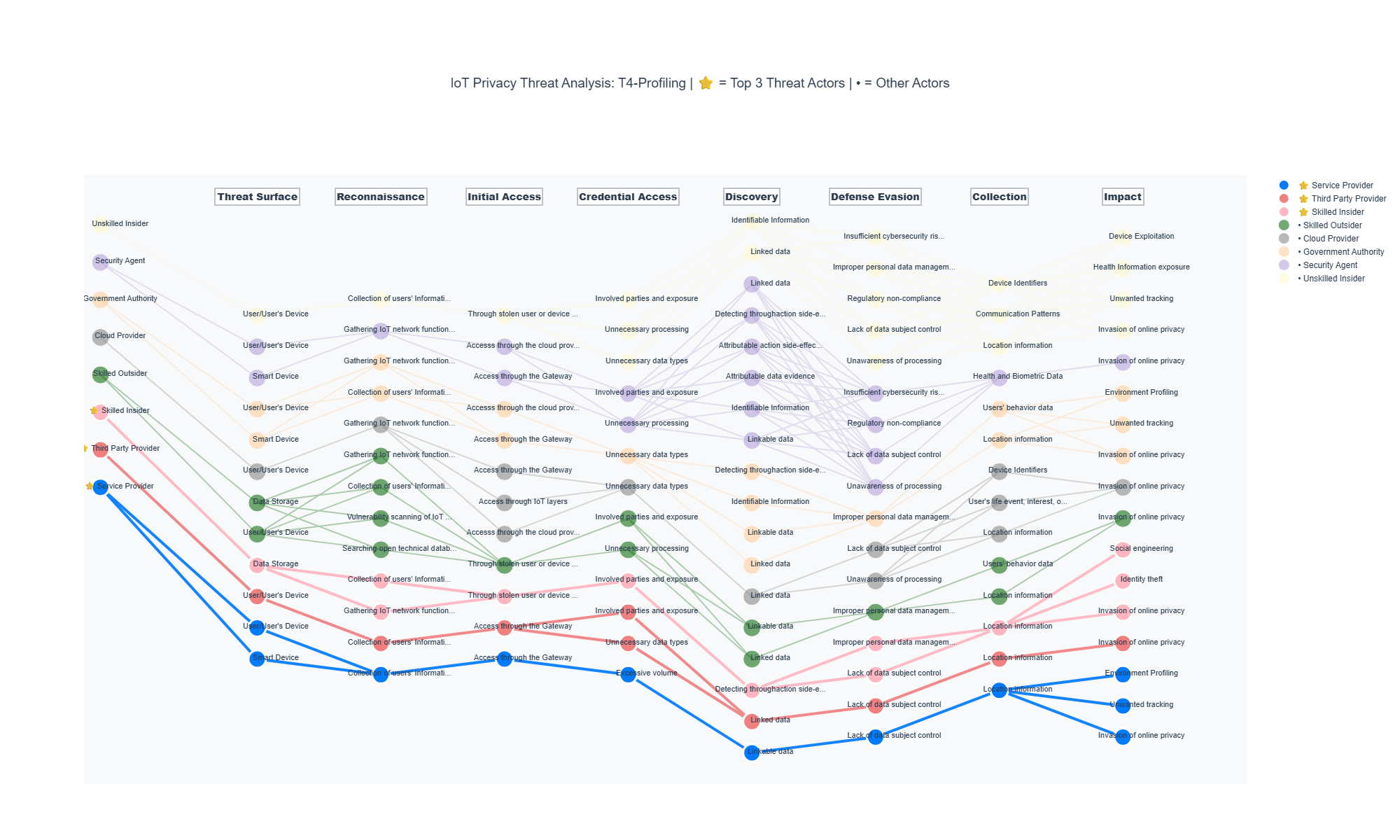}
        \caption{T4-Profiling network graph showing the Threat Actors techniques across all Threat Tactics}
        \label{fig:T4Net-graph1}
 \end{figure*}

 \textbf{Our observation}: The intentions of the top 3 threat actors that could be responsible for T4-Profiling differ, with the service provider and third-party provider having a common intention of invading the online privacy of the IoT user, mainly for business purposes through targeted marketing ads. Furthermore, third parties could track users' activities and locations, which could result in environmental profiling, restricting or directing certain services to the group or community to which the user belongs. However, skilled insider intention is malicious and considered dangerous, as the threat actors involved have already gained unauthorized access with the intention of invading the online privacy of IoT users. They also perform social engineering among IoT users within the same community, such as users of an IoT gadget that remotely monitors patient health. In some cases, they stole the identity of an IoT user to carry out more privacy threats, making it untraceable due to the authorized access and skill of this threat actor \cite{kinney2023analyzing}.

\subsection{T5-Impersonation}
\label{T5}
This occurs when the identity of an authorized user is stolen and used for an illegal purpose \cite{zainuddin2021study}.

  \begin{figure*}
        \centering
    \includegraphics[width=19cm,height=10cm]{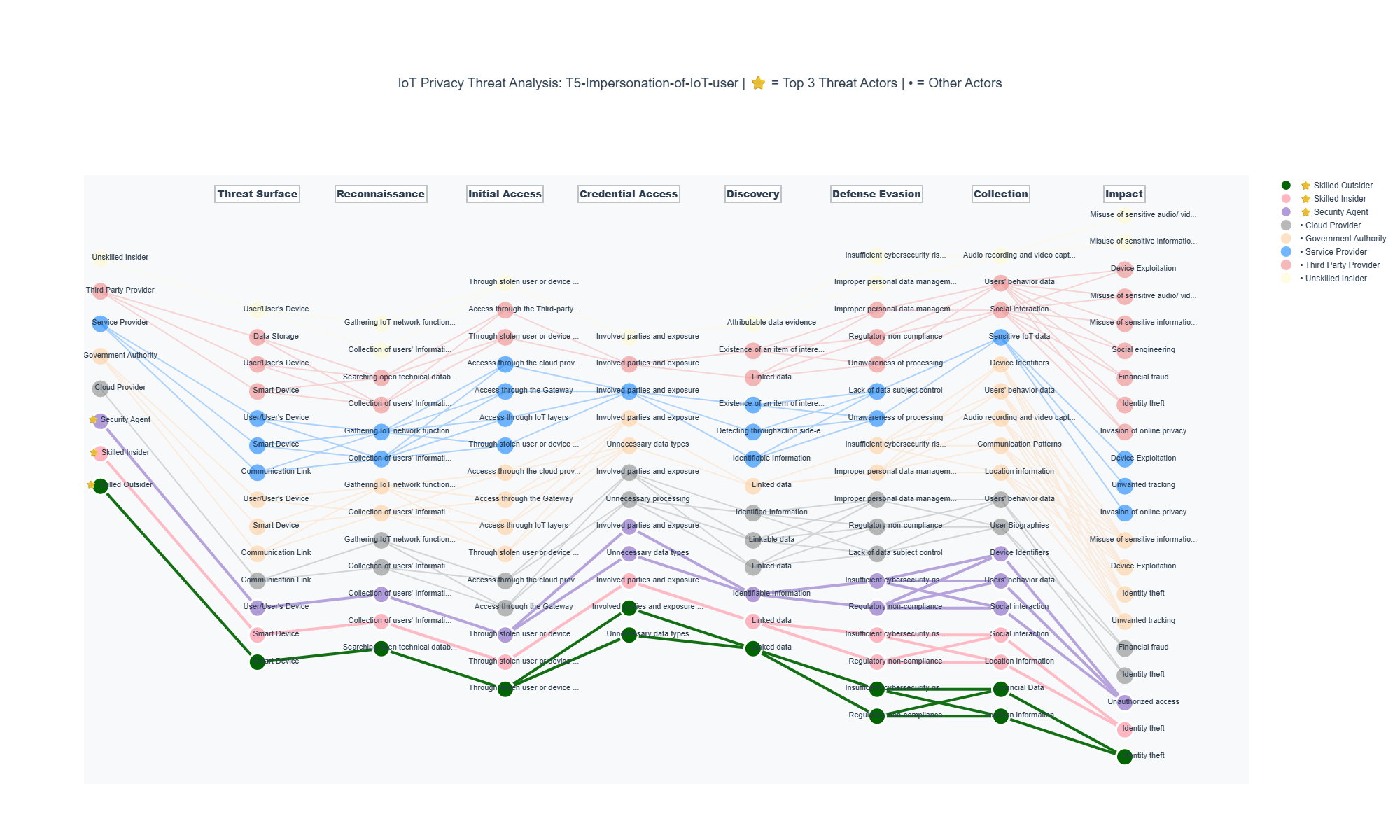}
        \caption{T5-Impersonation network graph showing the Threat Actors techniques across all Threat Tactics}
        \label{fig:T5Net-graph1}
 \end{figure*}

\textbf{Our observation:} In the T5 privacy threat, the intention of the skilled outsider and skilled insider is to steal the identity of the IoT user to carry out an impersonation privacy threat. This impersonation process could lead to further privacy threats, endangering not only other IoT users but also the device itself. However, a security agent could be responsible for the T5 privacy threat, with the intention of gaining access to IoT devices and user information, especially during the investigation process, which constitutes unauthorized access from the user's perspective. Despite using reconnaissance, initial access, credential access, and defense evasion techniques similar to those of other top 3 threat actors, the security agent targets the user/user device to impersonate a targeted user, primarily to lure other users connected to the user under investigation. This act is not malicious because the intention of the security agent is to have unauthorized access for the purpose of investigation. However, the activity of the skilled insider and outsider is malicious with the intention of stealing the identity of an IoT user or device.

\subsection{T6-Linkage (User or Device)}
\label{T6}
This threat involves revealing actual information about a user or device by connecting data from different sources  \cite{ziegeldorf2014privacy, ogonji2020survey, shaikh2019internet, al2016overview, jain2020privacy}.

  \begin{figure*}
        \centering
    \includegraphics[width=19cm,height=10cm]{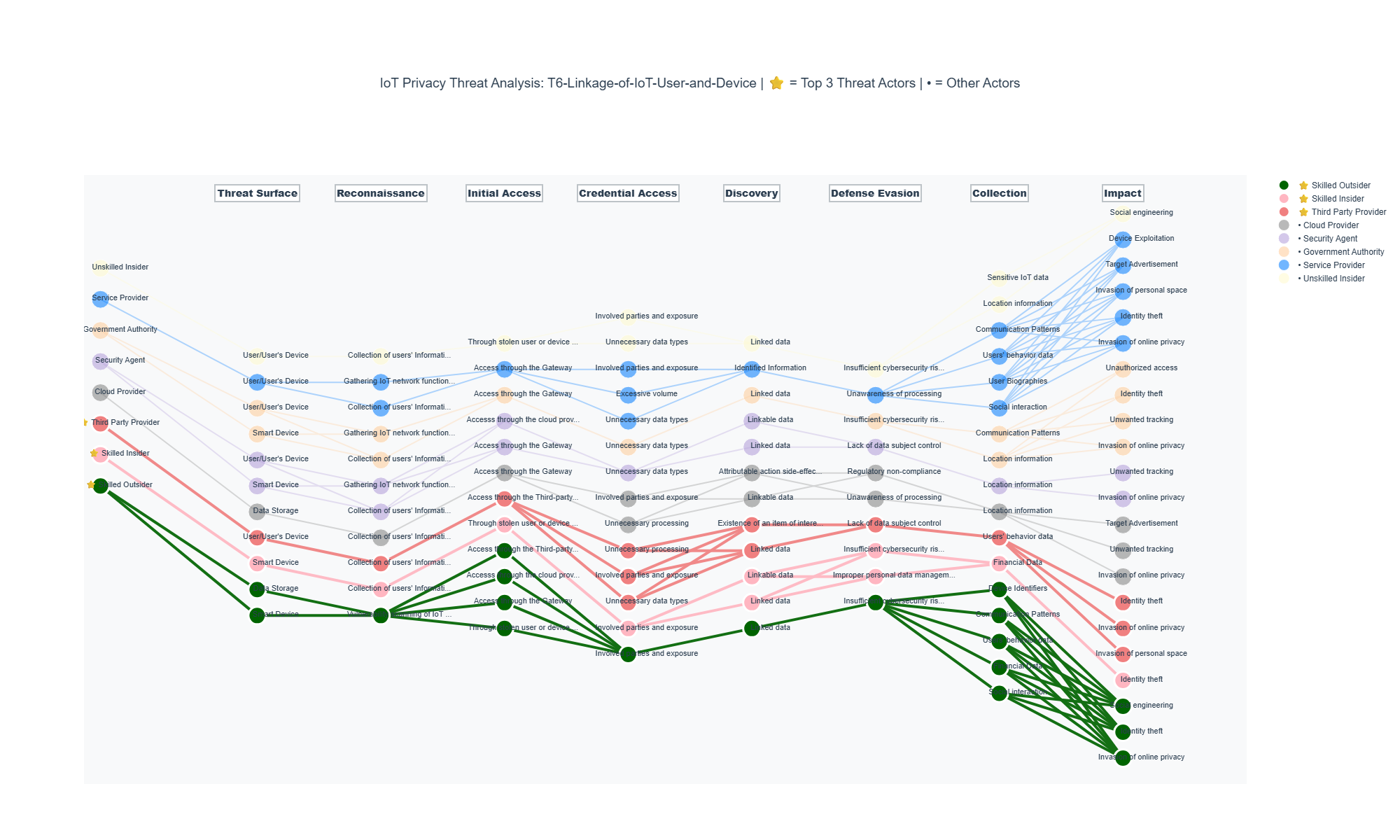}
        \caption{T6-Linkage (User or Device) network graph showing the Threat Actors techniques across all Threat Tactics}
        \label{fig:T6Net-graph1}
 \end{figure*}

 \textbf{Our observation}:  Similar to the T2 privacy threat, T6-Linkage of user or device data is most predominantly caused by a skilled outsider, a skilled insider, and a third-party provider. While the intention of both skilled outsiders and insiders is malicious, they exploit the identity of IoT devices and users by collecting sensitive data, such as financial information. Skilled outsiders could also invade the online privacy of IoT users and carry out social engineering through collected data, such as IoT user social interaction data, user behavior data, IoT device communication patterns, and device identifiers. The intention of third-party threat actors is not always malicious, as they tend to invade the personal space and online privacy of IoT users through the collected user behavior data. This enables third parties to present targeted ads to IoT users for marketing purposes.

\subsection{T7-Data Leakage}
\label{T7}
This threat results from the unaware exposure of sensitive information about a system, user, or device \cite{al2020overview} 

  \begin{figure*}
        \centering
    \includegraphics[width=19cm,height=11cm]{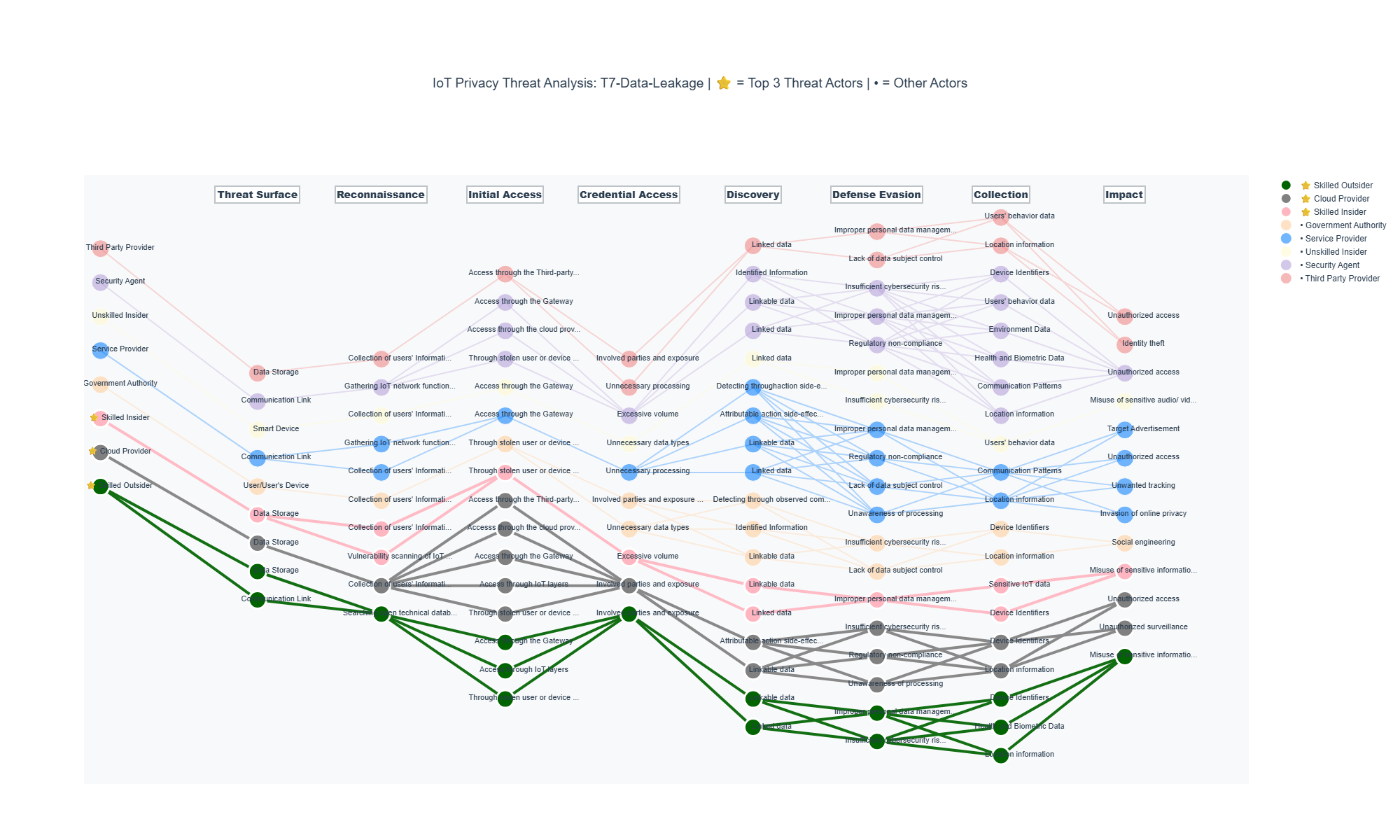}
        \caption{T7-Data Leakage network graph showing the Threat Actors techniques across all Threat Tactics}
        \label{fig:T7Net-graph1}
 \end{figure*}
 
 \textbf{Our observation: } The top 3 threat actors that could be responsible for T7-Data leakage are a skilled outsider, a cloud provider, and a skilled insider. The intention of both skilled outsiders and insiders is to exploit the sensitive information of IoT devices and user data collected to leak this information for financial gain. However, the intention of the cloud provider threat actor is to have unauthorized access and surveillance of the IoT device and the user. Some of this data can be shared with other third parties without the subject (i.e., IoT user or device) being aware of the data leakage to these parties. The intentions of skilled outsiders and insiders are malicious, as their ultimate goal is to exploit the IoT device and user data for financial gain or to launch further attacks on the IoT system. However, the cloud provider's intention is usually for business development and marketing purposes.

\subsection{T8-Jurisdiction Risk}
\label{T8}
Service providers' outsourcing of cloud-based applications and services to multiple parties has elevated the risk of disclosing sensitive information \cite{zainuddin2021study}.

  \begin{figure*}
        \centering
    \includegraphics[width=19cm,height=10cm]{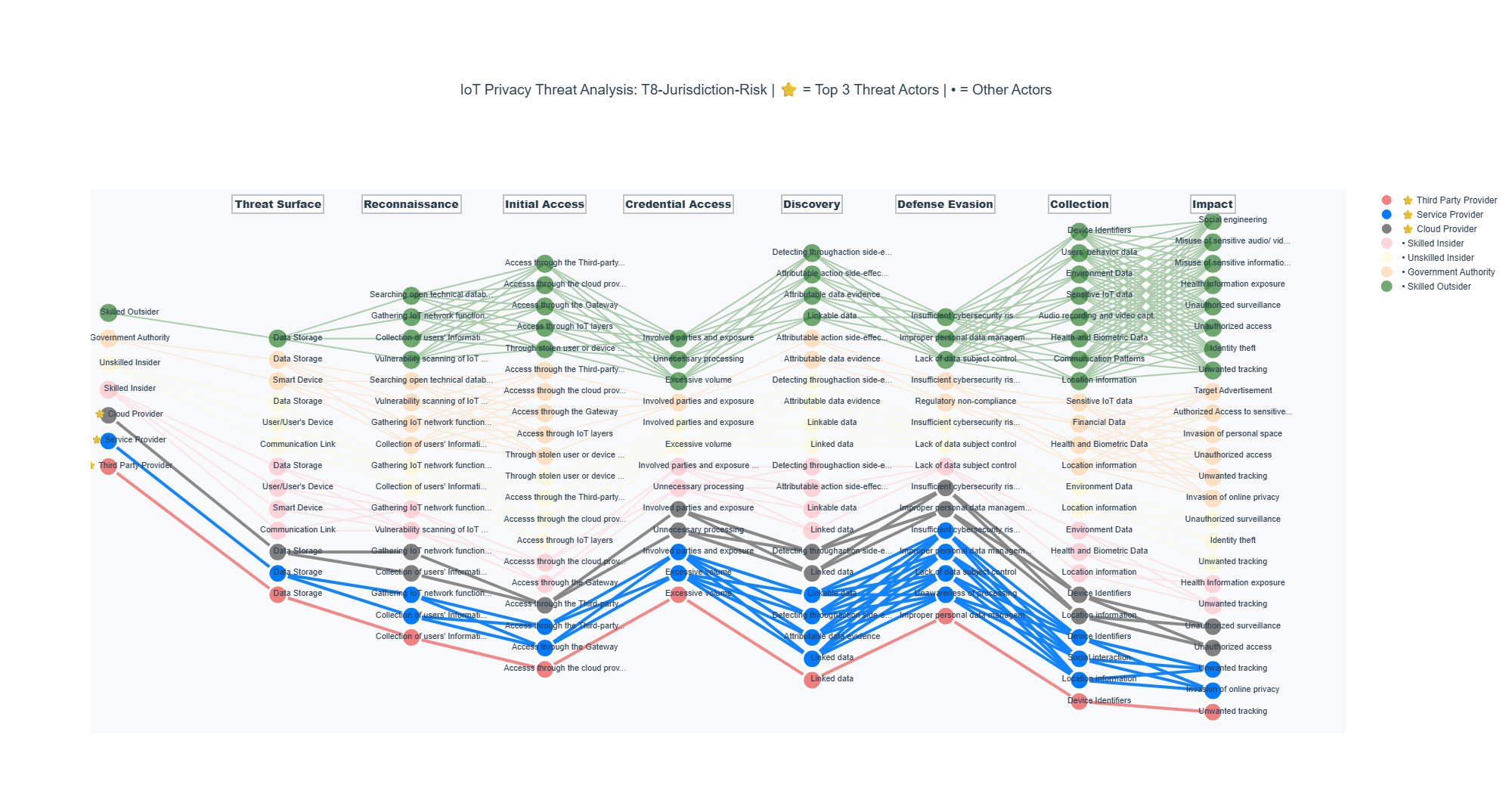}
        \caption{T8-Jurisdiction Risk network graph showing the Threat Actors techniques across all Threat Tactics}
        \label{fig:T8Net-graph1}
 \end{figure*}

\textbf{Our observation}: Although the top 3 privacy threats that could be responsible for T8-Jurisdiction risk are not malicious in their action, their action could endanger the IoT user and device indirectly. For instance, in a scenario where a data collector (service provider) shares data with the process (cloud provider) and third-party providers in order to provide quality service to IoT users. This data sharing involves no threat if the data is used for the purpose for which it was collected. However, any use of the data outside its original means of collection without the user's consent is a privacy threat to the user's data. This is an invasion of users' online privacy and a misuse of sensitive information. In some extreme cases, the collected data could be used for unwanted tracking and access for the purpose of targeted advertising, environmental profiling, or restricting or directing certain services to a group of people based on their common attribute.

\subsection{T9-Lifecycle transition}
\label{T9}
After IoT devices become obsolete and are no longer in use, they often retain sensitive information from their previous systems. Threat actors can use this information for malicious purposes  \cite{ziegeldorf2014privacy, ogonji2020survey, shaikh2019internet, alhalafi2019privacy, al2016overview, jain2020privacy}.

  \begin{figure*}
        \centering
    \includegraphics[width=19cm,height=10cm]{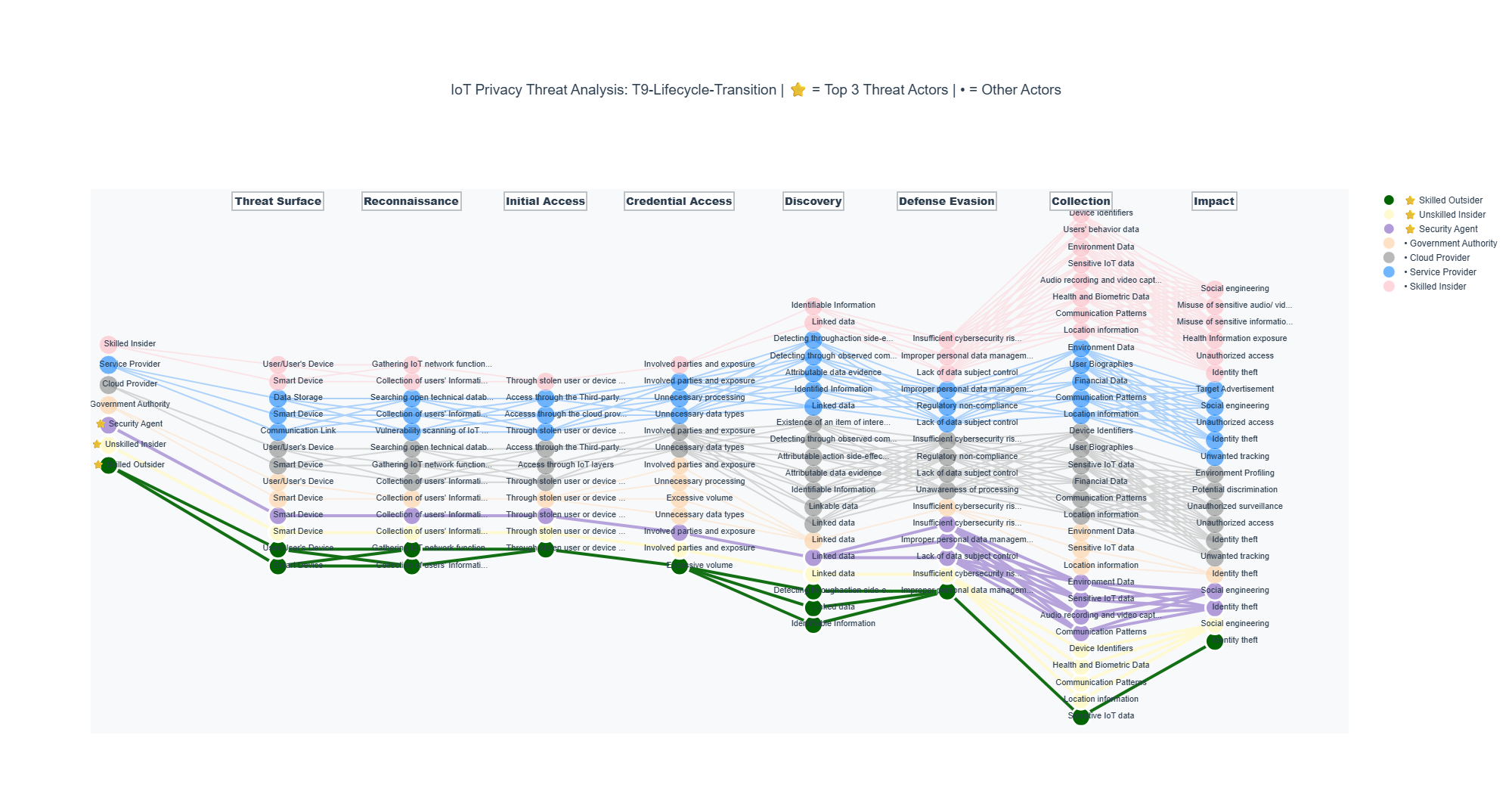}
        \caption{T9-Lifecycle transition network graph showing the Threat Actors techniques across all Threat Tactics}
        \label{fig:T9Net-graph1}
 \end{figure*}

\textbf{Our observation:} The top 3 threat actors usually responsible for the lifecycle transition of an IoT device are skilled outsiders, unskilled insiders, and security agents. Skilled outsiders intend to steal the identities of legitimate IoT users gathered from sensitive IoT data collected through vulnerabilities resulting from improper personal data management in the IoT system. However, an unskilled insider responsible for this T9 privacy threat can be categorized as an unintended insider according to \cite{kim2020review}, whose action of using stored data in decommissioned IoT device are non-malicious However, their action could cause harm or increases the probability of harm in the future to the IoT system accidental disclosure of sensitive data stored in these decommissioned IoT device. On the other hand, a security agent's intention could be categorized as non-malicious if it is for investigative purposes. Security agents often collect data, including audio recordings, video captures, communication patterns, and other sensitive IoT data, to investigate an individual or a group of individuals. The impact could be an act of identity theft of one of the target individual or a form of social engineering to gather more information for their investigation.

\subsection{T10-Inventory Attack}
\label{T10}
Illegitimate gathering of information about a subject's features, identity, and existence in an IoT system \cite{ziegeldorf2014privacy, ogonji2020survey,  alhalafi2019privacy}.

  \begin{figure*}
        \centering
    \includegraphics[width=19cm,height=10cm]{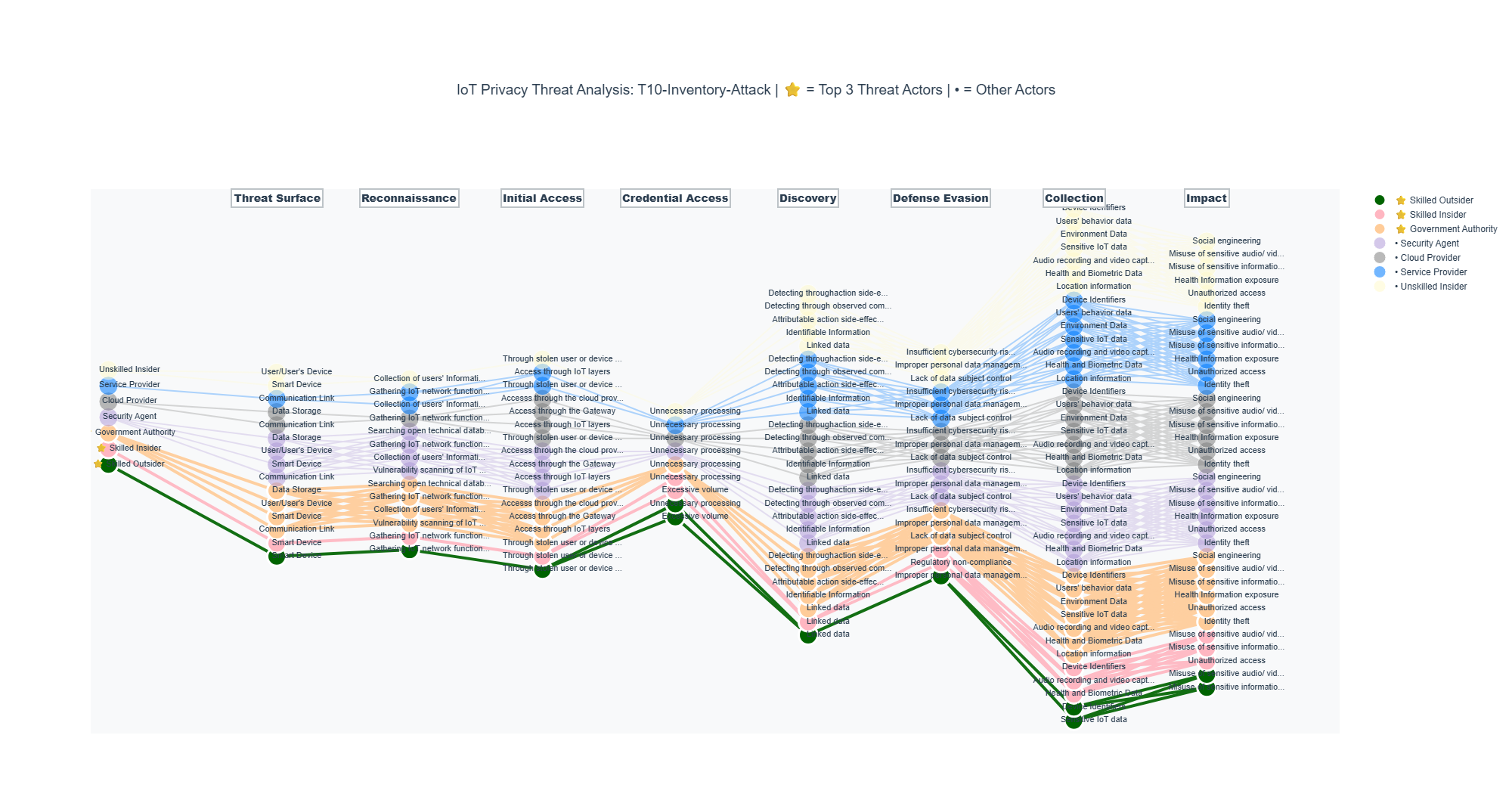}
        \caption{T10-Inventory Attack network graph showing the Threat Actors' techniques across all Threat Tactics}
        \label{fig:T10Net-graph1}
 \end{figure*}

\textbf{Our observation:} The skilled outsider, skilled insider, and government authority are the top 3 threat actors that could be responsible for the IoT device lifecycle transition privacy threat. Skilled outsider and skilled insider intentions are typically to misuse sensitive data and audio/video capture in IoT systems. A skilled insider could also use this data to gain unauthorized access into the IoT system through stolen administrative credentials obtained from linked data that could reveal these credentials. On the other hand, the government's intention is to collect details about IoT device identities and IoT user behaviors, as well as sensitive audio recordings and video captures from these obsolete devices, for the purpose of investigating their target individual or group for the purpose of social engineering \cite{konigs2022government}. The government authority can go as far as using the identity of one of their targets to lure others under investigation, or misuse sensitive data stored in these IoT devices to gain unauthorized access to the target individual or group being investigated, especially during an undercover operation.

\subsection{T11-Data Tampering}
\label{T11}
Ability of a threat actor to manipulate data after gaining unauthorized access to the data \cite{zainuddin2021study}.

  \begin{figure*}
        \centering
    \includegraphics[width=19cm,height=10cm]{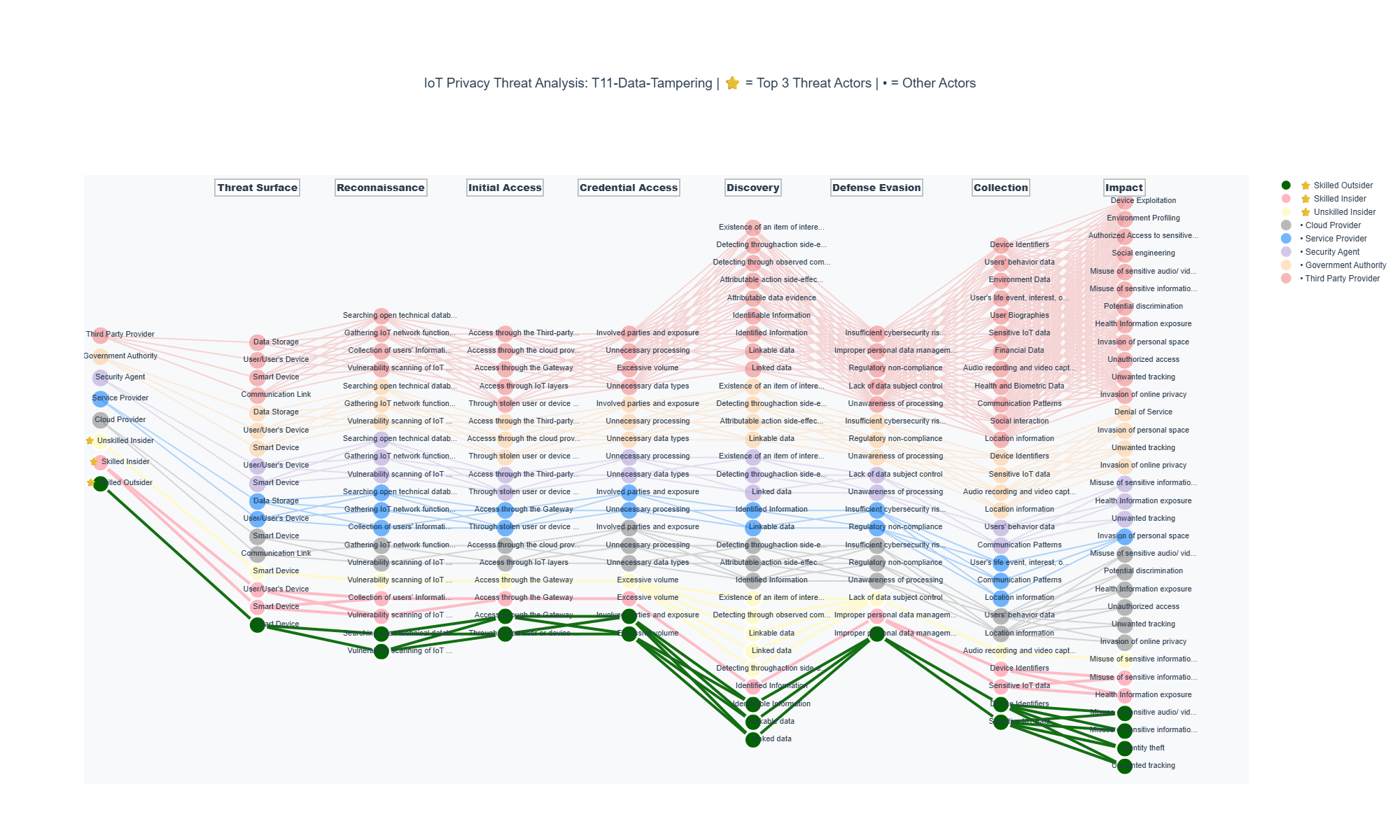}
        \caption{T11-Data Tampering network graph showing the Threat Actors' techniques across all Threat Tactics}
        \label{fig:T11Net-graph1}
 \end{figure*}
 \begin{figure*}
        \centering
    \includegraphics[width=19cm,height=10cm]{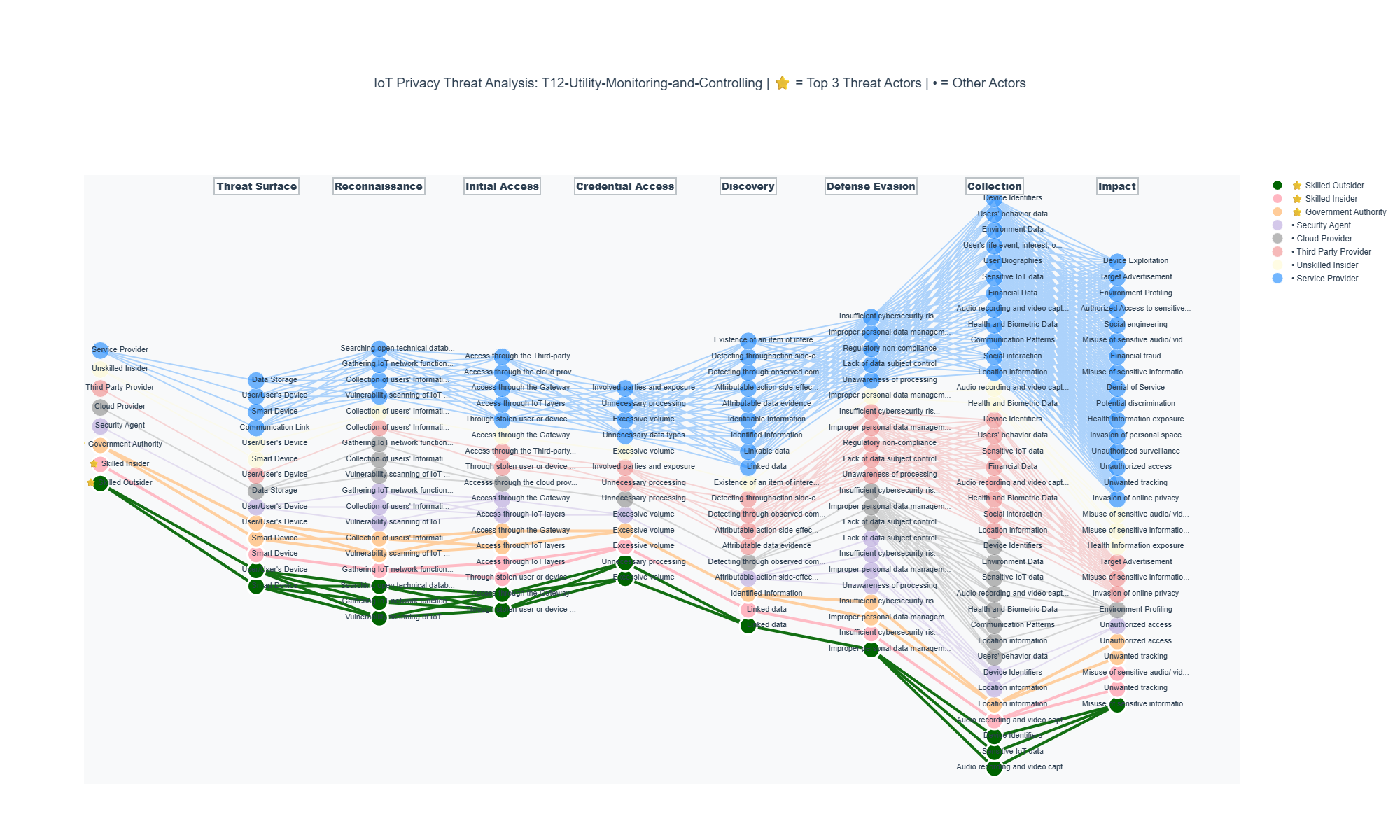}
        \caption{T12-Utility Monitoring and Controlling network graph showing the Threat Actors' techniques across all Threat Tactics}
        \label{fig:T12Net-graph1}
 \end{figure*}

\textbf{Our observation:}  Looking at the top 3 threat actors that could be responsible for the T11 privacy threat, a skilled outsider is the most prominent with the intention of misusing sensitive IoT device and user data, audio/video capture, as well as identity theft and tracking of IoT devices and users. A skilled insider has a similar intention to a skilled outsider. However, a skilled insider who is a traitor or masquerader \cite{kim2020review} could expose sensitive information, such as a patient's health information, in the case of the Internet of Medical Things (IoMT). Lastly, unskilled insider intentions are not usually malicious, especially if the unskilled insider's action or inaction is not with malicious intent. Misuse of information in these cases is usually a result of IoT user error, such as Misconfiguration or any other action out of ignorance. This action could reveal IoT-sensitive data, including audio recordings and video captures. Therefore, any of these threat actors' actions could directly or indirectly impact the IoT device and user, regardless of their intention.

\subsection{T12-Utility Monitoring and Controlling}
\label{T12}
Illegal remote monitoring or control of IoT system entities by threat actors \cite{seliem2018towards}.

\textbf{Our observation:} The top 3 threat actors for T12 privacy threat are skilled outsiders, skilled insiders, and government authorities, similar to the T10 privacy threat. The intention of a skilled outsider and a skilled insider is malicious due to the sensitive IoT data they could collect, which includes audio recordings and video captures. However, device identifiers such as IP addresses and MAC addresses can also be collected by the skilled outsider to gain more access to the IoT. The skilled insider tends to track IoT users or devices in an unauthorized manner. However, the government authority's action during T12 privacy threat is for surveillance and investigation purposes, granting them access to IoT user location information. This act results in IoT devices and users being tracked without their consent or granting unauthorized access to their data \cite{westerlund2021acceptance}.

\begin{landscape}

\label{appendixB}
\section*{Appendix B}

\begin{table}[ht]
\centering
\caption{Privacy threat table summarizing the top 3 threat actors, and their techniques during privacy threats T1-T12, including the collected data and the impact of their activities}
\fontsize{10pt}{10pt}\selectfont 
\begin{adjustbox}{max width=\linewidth}

\vspace{1.5cm}
\begin{tabular}{
      |L{1.6cm}|L{2.4cm}|L{2.3cm}|L{2.8cm}|L{2.3cm}|L{2.0cm}|L{2.0cm}|L{2.8cm}|L{2.4cm}|L{2.4cm}| }
\hline
\textbf{Privacy Threat} & 
\textbf{Prominent Threat Actor} & 
\textbf{Prominent Threat Surface} & 
\textbf{Reconnaissance} & 
\textbf{Initial Access} & 
\textbf{Credential Access} & 
\textbf{Discovery} & 
\textbf{Defense Evasion} & 
\textbf{Collection} & 
\textbf{Impact} \\
\hline
\multirow{3}{*}{T1} 
  & Cloud Provider 
  & Data storage 
  & Collection of users' information 
  & Access through the Gateway and  the cloud provider 
  & Unnecessary processing 
  & Linked data 
  & Improper personal data management 
  & Users' behavior data, Location Information 
  & Invasion of online privacy, Misuse of sensitive information \\
\cline{2-10}
  & Skilled Outsider 
  & Communication Link 
  & Collection of users' information, Vulnerability scanning of IoT network, Gathering of IoT network function, searching open technical database for digital certificates 
  & Through stolen user or device login credentials 
  & Unnecessary processing 
  & Linked data, identified information, Detection through trans(action) side effect 
  & Lack of data subject control 
  & Location Information 
  & Unauthorized access \\
\cline{2-10}
  & Service Provider 
  & User/User device 
  & Collection of users' information, Vulnerability scanning of IoT network 
  & Access through the Gateway and  the IoT layer 
  & Involved parties and exposure 
  & Linked data,  Detection through trans(action) side effect 
  & Lack of data subject control, Regulatory non-compliance, Improper personal data management 
  & Communication patterns, Users' behavior 
  & Invasion of online privacy, unwanted tracking \\
\hline
\multirow{3}{*}{T2} 
  & Third-Party provider 
  & Smart Device, User/user device 
  & Collection of users' information 
  & Through stolen user or device login credentials 
  & Involved parties and exposure 
  & Linked data,  Detection through trans(action) side effect 
  & Lack of data subject control, Regulatory non-compliance, Improper personal data management 
  & Location Information 
  & Financial Fraud \\
\cline{2-10}
  & Skilled Insider 
  & User/User device 
  & Collection of users' information 
  & Through stolen user or device login credentials 
  & Unnecessary processing 
  & Linked data
  & Improper personal data management, Insufficient cybersecurity risk management 
  & Location Information, Users' behavior data 
  & Misuse of sensitive information, Identity theft \\
\cline{2-10}
  & Skilled Outsider 
  & Smart Device, User/user device 
  & searching open technical database for digital certificates, Vulnerability scanning of IoT network 
  & Through stolen user or device login credentials 
  & Excessive volume, Unnecessary data types 
  & Linked Data 
  & Unawareness of processing data, Lack of data subject control, Improper personal data management 
  & Location Information 
  & Invasion of online privacy, Misuse of sensitive information, Financial fraud \\
\hline

\end{tabular}
\end{adjustbox}
\end{table}
\end{landscape}

\begin{landscape}
\begin{table}[ht]
\centering
\fontsize{10pt}{10pt}\selectfont 
\begin{adjustbox}{max width=\linewidth}
\begin{tabular}{
      |L{1.6cm}|L{2.4cm}|L{2.1cm}|L{2.8cm}|L{2.3cm}|L{2.0cm}|L{2.0cm}|L{2.8cm}|L{2.4cm}|L{2.4cm}| }
\hline
\textbf{Privacy Threat} & 
\textbf{Prominent Threat Actor} & 
\textbf{Prominent Threat Surface} & 
\textbf{Reconnaissance} & 
\textbf{Initial Access} & 
\textbf{Credential Access} & 
\textbf{Discovery} & 
\textbf{Defense Evasion} & 
\textbf{Collection} & 
\textbf{Impact} \\ 
\hline
\multirow{3}{*}{T3}
  & Skilled Outsider 
  & Smart Device 
  & Gathering of IoT network functions, Vulnerability scanning of IoT network 
  & Access through IoT layers 
  & Involved parties and exposure 
  & Linked data 
  & Regulatory non-compliance, Insufficient cybersecurity risk management 
  & Users' behavior data, Location Information 
  & Health Information exposure \\
\cline{2-10}
  & Security Agent 
  & Smart Device, User/user device 
  & Collection of users' information 
  & Access through IoT layers 
  & Unnecessary data types 
  & Linkable data, Attributable data evidence, Linked data 
  & Unawareness of processing data, Lack of data subject control, Improper personal data management, Insufficient cybersecurity risk management 
  & Device Identifiers, Location Information, Audio recording, and video capture 
  & Invasion of online privacy, unwanted tracking \\
\cline{2-10}
  & Third-Party provider 
  & User/User device 
  & Gathering of IoT network function, Collection of users' information 
  & Access through the Third-party provider 
  & Involved parties and exposure 
  & Linked data, Linkable data 
  & Insufficient cybersecurity risk management, Regulatory non-compliance 
  & Location Information 
  & Invasion of online privacy \\
\hline

\multirow{3}{*}{T4} 
  & Service Provider 
  & Smart Device, User/user device 
  & Collection of users' information 
  & Access through the gateway 
  & Excessive volume 
  & Linkable data 
  & Lack of data subject control 
  & Location Information 
  & Invasion of online privacy, Unwanted tracking, Environment Profiling \\
\cline{2-10}
  & Third-Party provider 
  & User/User device 
  & Collection of users' information 
  & Access through the gateway 
  & Unnecessary data types, Involved parties and exposure 
  & Linked data 
  & Lack of data subject control 
  & Location Information 
  & Invasion of online privacy \\
\cline{2-10}
  & Skilled Insider 
  & Data storage 
  & Collection of users' information, Gathering IoT network functions 
  & Through stolen user or device login credentials 
  & Involved parties and exposure 
  & Detection through trans(action) side effect 
  & Lack of data subject control, Improper personal data management 
  & Location Information 
  & Invasion of online privacy, Identity theft, Social engineering \\
\hline
\end{tabular}
\end{adjustbox}
\end{table}
\end{landscape}

\begin{landscape}
\begin{table}[ht]
\centering
\fontsize{10pt}{10pt}\selectfont 
\begin{adjustbox}{max width=\linewidth}
\begin{tabular}{
      |L{1.6cm}|L{2.4cm}|L{2.1cm}|L{2.8cm}|L{2.3cm}|L{2.0cm}|L{2.0cm}|L{2.8cm}|L{2.4cm}|L{2.4cm}| }
\hline
\textbf{Privacy Threat} & 
\textbf{Prominent Threat Actor} & 
\textbf{Prominent Threat Surface} & 
\textbf{Reconnaissance} & 
\textbf{Initial Access} & 
\textbf{Credential Access} & 
\textbf{Discovery} & 
\textbf{Defense Evasion} & 
\textbf{Collection} & 
\textbf{Impact} \\
\hline
\multirow{3}{*}{T5} 
  & Skilled Outsider 
  & Smart Device 
  & searching open technical database for digital certificates 
  & Through stolen user or device login credentials 
  & Unnecessary data types, Involved parties and exposure 
  & Linked data 
  & Regulatory non-compliance, Insufficient cybersecurity risk management 
  & Financial Data, Location Information 
  & Identity theft \\
\cline{2-10}
  & Skilled Insider 
  & Smart Device 
  & Collection of users' information 
  & Through stolen user or device login credentials 
  & Involved parties and exposure 
  & Linked data 
  & Regulatory non-compliance, Insufficient cybersecurity risk management 
  & Social Interaction, Location Information 
  & Identity theft \\
\cline{2-10}
  & Security Agent 
  & User/User device 
  & Collection of users' information 
  & Through stolen user or device login credentials 
  & Involved parties and exposure, Unnecessary data types 
  & Identifiable Information 
  & Regulatory non-compliance, Insufficient cybersecurity risk management 
  & Social Interaction, Users' behavior data, Device Identifiers 
  & Unauthorized access \\
\hline
\multirow{3}{*}{T6} 
  & Skilled Outsider 
  & Smart Device, Data Storage 
  & Vulnerability scanning of IoT network 
  & Through stolen user or device login credentials, Access through Gateway, Access through Cloud provider, Access through Third-party 
  & Involved parties and exposure, Unnecessary data types 
  & Linked data 
  & Insufficient cybersecurity risk management 
  & Social Interaction, Financial Data, Users' behavior data, Communication patterns, Device Identifiers 
  & Invasion of online privacy, Identity theft, Social engineering \\
\cline{2-10}
  & Skilled Insider 
  & Smart Device 
  & Collection of users' information 
  & Through stolen user or device login credentials 
  & Involved parties and exposure, Unnecessary data types 
  & Linked data, Linkable data 
  & Insufficient cybersecurity risk management, Improper personal data management 
  & Financial Data 
  & Identity theft \\
\cline{2-10}
  & Third-Party provider 
  & User/User device 
  & Collection of users' information 
  & Access through Third-party 
  & Unnecessary processing, Involved parties and exposure, Unnecessary data types 
  & Linked data, Existence of an item of interest revealed through system response 
  & Lack of data subject control 
  & Users' behavior data 
  & Invasion of personal space, Invasion of online privacy, Identity theft \\
\hline
\end{tabular}
\end{adjustbox}
\end{table}
\end{landscape}

\begin{landscape}

\begin{table}[ht]
\centering
\fontsize{10pt}{10pt}\selectfont 
\begin{adjustbox}{max width=\linewidth}
\begin{tabular}{
      |L{1.6cm}|L{2.4cm}|L{2.1cm}|L{2.8cm}|L{2.3cm}|L{2.0cm}|L{2.0cm}|L{2.8cm}|L{2.4cm}|L{2.4cm}| }
\hline
\textbf{Privacy Threat} & 
\textbf{Prominent Threat Actor} & 
\textbf{Prominent Threat Surface} & 
\textbf{Reconnaissance} & 
\textbf{Initial Access} & 
\textbf{Credential Access} & 
\textbf{Discovery} & 
\textbf{Defense Evasion} & 
\textbf{Collection} & 
\textbf{Impact} \\
\hline
\multirow{3}{*}{T7} 
  & Skilled Outsider 
  & Communication Link, Data storage 
  & searching open technical database for digital certificates 
  & Through stolen user or device login credentials, Access through IoT layers, Gateway 
  & Involved parties and exposure, Unnecessary data types 
  & Linked data, Linkable data 
  & Insufficient cybersecurity risk management, Improper personal data management 
  & Location Information, Health and Biometric Data, Device Identifiers 
  & Misuse of sensitive information \\
\cline{2-10}
  & Cloud Provider 
  & Data storage 
  & Collection of users' information 
  & Through stolen user or device login credentials, Access through IoT layers, Gateway, Cloud provider, Third-party 
  & Involved parties and exposure, Unnecessary data types 
  & Linkable, Attributable action side-effect evidence 
  & Insufficient cybersecurity risk management, Regulatory non-compliance, Unawareness of processing data 
  & Location Information, Device Identifiers 
  & Unauthorized access, Unauthorized Surveillance \\
\cline{2-10}
  & Skilled Insider 
  & Data storage 
  & Collection of users' information, Vulnerability scanning of IoT network 
  & Through stolen user or device login credentials 
  & Excessive volume 
  & Linked data, Linkable data 
  & Improper personal data management 
  & Device Identifiers, Sensitive IoT data 
  & Misuse of sensitive information \\
\hline
\multirow{3}{*}{T8} 
  & Third-Party provider 
  & Data storage 
  & Collection of users' information 
  & Access through Cloud provider 
  & Excessive volume 
  & Linked data 
  & Improper personal data management 
  & Device Identifiers 
  & Unwanted tracking \\
\cline{2-10}
  & Service Provider 
  & Data storage 
  & Collection of users' information, Gathering IoT network functions 
  & Access through Gateway, Third-party 
  & Excessive volume, Involved parties, and exposure 
  & Linked data, Attributable data evidence, Detection through trans(action) side effect, data, Linkable data 
  & Unawareness of processing data, Lack of data subject control, Improper personal data management, Insufficient cybersecurity risk management 
  & Location Information, Social Interaction, Device Identifiers 
  & Invasion of online privacy, Unwanted tracking \\
\cline{2-10}
  & Cloud Provider 
  & Data storage 
  & Collection of users' information, Gathering IoT network functions 
  & Access through Third-party 
  & Unnecessary processing, Involved parties, and exposure 
  & Linked data, Detection through trans(action) side effect 
  & Insufficient cybersecurity risk management, Improper personal data management 
  & Location Information, Device Identifiers 
  & Unauthorized access, Unauthorized Surveillance \\
\hline

\end{tabular}
\end{adjustbox}
\end{table}
\end{landscape}

\begin{landscape}
\begin{table}[ht]
\centering
\fontsize{9.5pt}{9.5pt}\selectfont 
\begin{adjustbox}{max width=\linewidth}
\begin{tabular}{
      |L{1.6cm}|L{2.4cm}|L{2.1cm}|L{2.8cm}|L{2.3cm}|L{2.0cm}|L{2.0cm}|L{2.8cm}|L{2.4cm}|L{2.4cm}| }
\hline
\textbf{Privacy Threat} & 
\textbf{Prominent Threat Actor} & 
\textbf{Prominent Threat Surface} & 
\textbf{Reconnaissance} & 
\textbf{Initial Access} & 
\textbf{Credential Access} & 
\textbf{Discovery} & 
\textbf{Defense Evasion} & 
\textbf{Collection} & 
\textbf{Impact} \\
\hline
\multirow{3}{*}{T9} 
  & Skilled Outsider 
  & Smart Device, User/user device 
  & Collection of users' information, Gathering IoT network functions 
  & Through stolen user or device login credentials 
  & Excessive volume 
  & Identifiable Information, Linked data, Detection through trans(action) side effect 
  & Improper personal data management 
  & Sensitive IoT data 
  & Identity theft \\
\cline{2-10}
  & Unskilled Insider 
  & Smart Device 
  & Collection of users' information 
  & Through stolen user or device login credentials 
  & Involved parties and exposure 
  & Linked data 
  & Insufficient cybersecurity risk management 
  & Location Information, Communication patterns, Health and Biometric Data, Device Identifiers 
  & Social engineering \\
\cline{2-10}
  & Security Agent 
  & Smart Device 
  & Collection of users' information 
  & Through stolen user or device login credentials 
  & Involved parties and exposure 
  & Linked data 
  & Insufficient cybersecurity risk management, Improper personal data management, Lack of data subject control 
  & Environment Data, Sensitive IoT data, Audio recording and video capture, Communication patterns 
  & Identity theft, Social engineering \\
\hline
\multirow{3}{*}{T10} 
  & Skilled Outsider 
  & Smart Device 
  & Gathering IoT network functions 
  & Through stolen user or device login credentials 
  & Excessive volume, Unnecessary processing 
  & Linked data 
  & Improper personal data management 
  & Sensitive IoT data, Device Identifiers 
  & Misuse of sensitive information, Misuse of sensitive audio/video data \\
\cline{2-10}
  & Skilled Insider 
  & Smart Device 
  & Gathering IoT network functions 
  & Through stolen user or device login credentials 
  & Unnecessary processing, Excessive volume 
  & Linked data 
  & Regulatory non-compliance, Improper personal data management 
  & Device Identifiers, Audio recording and video capture, Health and Biometric Data 
  & Misuse of sensitive audio/video data, Misuse of sensitive information, Unauthorized access \\
\cline{2-10}
  & Government Authority 
  & Data storage, User/User device, Smart Device, Communication Link 
  & Searching open technical database for digital certificates, Gathering IoT network function, Collection of users' information, Vulnerability scanning of IoT network 
  & Access through Cloud provider, Gateway, IoT layers, Through stolen user or device login credentials 
  & Unnecessary processing 
  & Linked data, Identifiable Information, Attributable action side-effect evidence, Detection through observed communication, Detection through trans(action) side effect 
  & Insufficient cybersecurity risk management, Improper personal data management, Lack of data subject control 
  & Device Identifiers, Users' behavior data, Environment Data, Sensitive IoT data, Audio recording and video capture, Health and Biometric Data, Location Information 
  & Social engineering, Misuse of sensitive audio/video data, Misuse of sensitive information, Identity theft, Unauthorized access \\
\hline
\end{tabular}
\end{adjustbox}
\end{table}
\end{landscape}

\begin{landscape}
\begin{table}[ht]
\centering
\fontsize{9.5pt}{9.5pt}\selectfont 
\begin{adjustbox}{max width=\linewidth}
\begin{tabular}{
      |L{1.6cm}|L{2.4cm}|L{2.1cm}|L{2.8cm}|L{2.3cm}|L{2.0cm}|L{2.0cm}|L{2.8cm}|L{2.4cm}|L{2.4cm}| }
\hline
\textbf{Privacy Threat} & 
\textbf{Prominent Threat Actor} & 
\textbf{Prominent Threat Surface} & 
\textbf{Reconnaissance} & 
\textbf{Initial Access} & 
\textbf{Credential Access} & 
\textbf{Discovery} & 
\textbf{Defense Evasion} & 
\textbf{Collection} & 
\textbf{Impact} \\
\hline

\multirow{3}{*}{T11} 
  & Skilled Outsider 
  & Smart Device 
  & Vulnerability scanning of IoT network, searching open technical database for digital certificates 
  & Through stolen user or device login credentials, Access through Gateway 
  & Excessive volume, Involved parties and exposure 
  & Linked data, Linkable data, Identifiable Information 
  & Improper personal data management 
  & Sensitive IoT data, Device Identifiers 
  & Misuse of sensitive information, Misuse of sensitive audio/video data, Identity theft, Unwanted tracking \\
\cline{2-10}
  & Skilled Insider 
  & Smart Device, User/user device 
  & Collection of users' information, Vulnerability scanning of IoT network 
  & Access through Gateway 
  & Excessive volume 
  & Identifiable Information 
  & Improper personal data management 
  & Device Identifiers, Sensitive IoT data 
  & Misuse of sensitive information, Health Information exposure \\
\cline{2-10}
  & Unskilled Insider 
  & Smart Device 
  & Vulnerability scanning of IoT network 
  & Access through Gateway 
  & Excessive volume 
  & Linked data, Linkable data, Detection through observed communication, Existence of an item of interference 
  & Lack of data subject control 
  & Audio recording and video capture 
  & Misuse of sensitive information \\
\hline
\multirow{3}{*}{T12} 
  & Skilled Outsider 
  & Smart Device, User/user device 
  & Searching open technical database for digital certificates, Gathering of IoT network function, Vulnerability scanning of IoT network 
  & Through stolen user or device login credentials, Access through Gateway 
  & Excessive volume, Unnecessary processing 
  & Linked data 
  & Improper personal data management 
  & Device Identifiers, Sensitive IoT data, Audio recording and video capture 
  & Misuse of sensitive information \\
\cline{2-10}
  & Skilled Insider 
  & Smart Device 
  & Gathering of IoT network function 
  & Access through IoT layers, Through stolen user or device login credentials 
  & Excessive volume 
  & Linked data 
  & Insufficient cybersecurity risk management 
  & Audio recording and video capture 
  & Unwanted tracking, Misuse of sensitive audio/video data \\
\cline{2-10}
  & Government Authority 
  & User/user device, Smart Device 
  & Collection of users' information, Vulnerability scanning of IoT network 
  & Access through Gateway, IoT layers 
  & Excessive volume 
  & Identifiable Information 
  & Insufficient cybersecurity risk management, Improper personal data management 
  & Location Information 
  & Unauthorized access, Unwanted tracking \\

\hline
\end{tabular}
\end{adjustbox}
\end{table}
\end{landscape}

 \end{document}